\documentclass[fleqn,usenatbib]{mnras}
\usepackage{newtxtext,newtxmath}
\usepackage[T1]{fontenc}
\DeclareRobustCommand{\VAN}[3]{#2}
\let\VANthebibliography\thebibliography
\def\thebibliography{\DeclareRobustCommand{\VAN}[3]{##3}\VANthebibliography}
\usepackage{graphicx}	
\usepackage{amsmath}	

\usepackage{hyperref}
\usepackage{chngcntr}
\usepackage{ulem}

\usepackage{ulem}

\title[Minimum initial mass for CCSNe at high Z]{Bridging the Gap between Intermediate and Massive Stars II: $M_\text{mas}$ for the most metal--rich stars and implications for Fe CCSNe rates}

\author[Cinquegrana, Joyce and Karakas]{
Giulia C. Cinquegrana,$^{1,2}$\thanks{E-mail: giulia.cinquegrana1@monash.edu}
Meridith Joyce,$^{3,4,5}$
Amanda I. Karakas$^{1,2}$
\\
$^{1}$School of Physics \& Astronomy, Monash University, Clayton VIC 3800, Australia\\
$^{2}$ARC Centre of Excellence for All Sky Astrophysics in 3 Dimensions (ASTRO 3D) \\
$^{3}$Konkoly Observatory, Research Centre for Astronomy and Earth Sciences, H-1121 Budapest Konkoly Th. M. \'ut 15-17., Hungary\\
$^{4}$CSFK, MTA Centre of Excellence, Budapest, Konkoly Thege Mikl\'os \'ut 15-17., H-1121, Hungary\\
$^{5}$Lasker Fellow, Space Telescope Science Institute, 3700 San Martin Drive, Baltimore, MD 21218, USA
}

\date{Accepted XXX. Received YYY; in original form ZZZ}

\pubyear{2023}

\begin{document}
\label{firstpage}
\pagerange{\pageref{firstpage}--\pageref{lastpage}}
\maketitle

\begin{abstract}

The minimum initial mass required for a star to explode as an Fe core collapse supernova, typically denoted $M_\text{mas}$, is an important quantity in stellar evolution because it defines the border between intermediate mass and massive stellar evolutionary paths. The precise value of $M_\text{mas}$ carries implications for models of galactic chemical evolution and the calculation of star formation rates. Despite the fact that stars with super solar metallicities are commonplace within spiral and some giant elliptical galaxies, there are currently no studies of this mass threshold in super metal-rich models with $Z>0.05$. Here, we study the minimum mass necessary for a star to undergo an Fe core collapse supernova when its initial metal content falls in the range $2.5\times 10^{-3} \leq Z \leq 0.10$. Although an increase in initial $Z$ corresponds to an increase in the Fe ignition threshold for $Z \approx 1\times 10^{-3}$ to $Z\approx0.04$, we find that there is a steady reversal in trend that occurs for $Z > 0.05$. Our super metal-rich models thus undergo Fe core collapse at \textit{lower} initial masses than those required at solar metallicity. Our results indicate that metallicity--dependent curves extending to $Z=0.10$ for the minimum Fe ignition mass should be utilised in galactic chemical evolution simulations to accurately model supernovae rates as a function of metallicity, particularly for simulations of metal-rich spiral and elliptical galaxies. 
\end{abstract}

\begin{keywords}
keyword1 -- keyword2 -- keyword3
\end{keywords}



\section{Introduction} \label{sec:intro}

The bulges of spiral galaxies, such as the Milky Way, are known to harbour metal-rich stellar populations \citep{mcwilliam1994first, Lepine11, Feltzing13, Do15, ryde2015chemical, Bensby2017, Joyce2022}, as are giant elliptical galaxies (M49; \citealt{Cohen03}). Here, the term `metal-rich' is used to describe a metallicity greater than the sun, i.e., $Z \gtrsim Z_{\odot}$. This work is motivated by candidates residing on the extreme end of the metal-rich range; examples of regions with an observed metallicity ([M/H], [Fe/H] or [Z/H]) greater than $\sim 0.4$ include the Globular Clusters (GCs) in M49 (NGC 4472) from \citet{Cohen03}, which extend to $\rm [Z/H] = 0.98$. As part of the Fornax 3D project \citep{fornax3d}, \citet{fahrion2020fornax} derive metallicities for 187 GCs that span 23 galaxies (comprising the Fornax cluster). They calculate $\approx$15 GCs with $0 \leq \rm [M/H] \leq 0.5$. Resolved metal-rich stars (with an $\rm [Fe/H]$ up to 0.55) within our galaxy have also been captured with the HARPS (High-Accuracy Radial velocity Planetary Searcher) GTO planet search program \citep{phase2003setting, curto2010harps, santos2011harps} in \citet{mena2017chemical}. Evolved metal-rich stars can be found in the open cluster NGC6791; spectroscopic metallicities are reported for these stars between $0.4 \leq \rm [M/H] \leq 0.45$ in the second APOKASC catalogue \citep{pinsonneault2018second}. Additional stars with [Fe/H] > 0.4 were found by \citet{thorsbro2020detailed} in the Galactic centre, which spans $-3 \leq \rm [Fe/H] \leq 1$ \citep{ness2016metallicity}. \citet{do2015discovery} measures some of the most enriched stars in the Galactic centre, with [M/H] up to 0.96.

Characterizing the behaviour of these metal-rich stars requires stellar evolution models with super-solar initial abundances. Such models were presented in \citet{Karakas22paper1} and \citet{Cinquegrana22paper2}, which focused on the peculiar evolution and nucleosynthetic behaviour of low and intermediate mass super metal-rich stars with $Z_{\rm max}=0.10$ (e.g., those that will end their lives as CO, hybrid CO-ONe or ONe white dwarfs). Other existing sets of metal-rich models for the low and intermediate mass regime include \citet{bono1997evolutionary, Siess07, Weiss09, Karakas14He, Karakas16, Ventura20} with $Z_{\rm max} \leq 0.04$ and \citet{Fagotto1994evolutionary, valcarce13, Marigo13, Marigo17} with $0.04 < Z_{\rm max} \leq 0.10$. For models that span the entire mass range, see \citet{bono00} ($Z_{\rm max} \leq 0.04$) and \citet{mowlavi1998grids, Mowlavi98, salasnich00, Meynet06, claret07, Bertelli08, Choi16} ($0.04 < Z_{\rm max}  \leq 0.10$).

A key finding of the models from \citet{Karakas22paper1} was the non-monotonic behaviour of central C ignition with increasing metallicity. Namely, our most metal-rich models ignite core C burning at \textit{lower} initial masses than required at solar metallicity. In this work, we extend our initial analysis to determine whether this behaviour is mirrored for higher order burning stages and end--of--life scenarios. In particular, we identify the initial mass required for stars to undergo an Fe core collapse supernova (hereafter, Fe CCSNe) as a function of metallicity. Following convention, we denote this quantity $M_{\rm mas}$.

The initial mass function (IMF) of galaxies skews heavily towards lower stellar masses and are thus well constrained in this regime. Given that the higher--mass end of the IMF has a much smaller number of calibrators, a small decrease in $M_{\rm mas}$ would correspond to a significant increase in the number of supernovae (SNe) events predicted to occur in a given region. The nucleosynthetic contributions from low and intermediate mass stars differ greatly from the explosive nucleosynthesis products generated by massive stars, thus defining an accurate SNe rate is vital for galactic chemical evolution models. This is sharply demonstrated by \citet{pillepich2018simulating}, who increase $M_{\rm mas}$ from 6\(M_\odot\) in the original Illustris models to 8\(M_\odot\) for the IlustrisTNG models in their current work \citep{vogelsberger2014introducing, vogelsberger2014properties, genel2014introducing, sijacki2015illustris}. This change results in 30\% fewer SNe events and so has a substantial downstream impact on the [Mg/Fe] yield \citep{naiman2018first}, given type II SNe are significant producers of Mg \citep[e.g.][]{Kobayashi20}. Likewise, given the short lifetimes of massive stars, $M_{\rm mas}$ is often used to predict rates of star formation \citep{keane2008birthrates, botticella2012comparison}. 

The $M_{\rm mas}$ quantity has been investigated in a stellar modelling context by a number of groups. At solar metallicity, \citet{Woosley15remarkable} and \citet{jones2013advanced} derive values of $M_{\rm mas}=9$ and 9.5\(M_\odot\) using KEPLER \citep{Weaver78presupernova, woosley2002} and \texttt{MESA} (\citealt{Paxton10instrument1, Paxton13instrument2, Paxton15instrument3, Paxton18instrument4, Paxton19instrument5, Jermyn22instrument6}). \citet{Poelarends08supernova} consider the same initial composition but with three different stellar evolution codes: STERN \citep{langer1998coupled, heger2000presupernova}, KEPLER and EVOL (\citealt{Blocker95a, Herwig04dredge, Herwig04nuclear}). They find $M_{\rm mas}=13$\(M_\odot\), 9.2\(M_\odot\) and 10.5\(M_\odot\), respectively. The models of \citet{bressan1993evolutionary} are calculated with a slightly super-solar composition, $Z=0.02$, as opposed to $Z=0.015$ in \citet{Woosley15remarkable}. They find the distinction between intermediate mass and massive stars falls in the $5-6$\(M_\odot\) range. 

The following studies map $M_{\rm mas}$ over a $Z$ range: \citet{Eldridge04prog} using Cambridge STARS \citep{Eggleton71evolution} to cover $Z=1\times 10^{-5}$ to $Z=0.05$. \citet{Doherty15super} calculate $M_{\rm mas}$ for $Z=0.0001$ to $0.02$ using MONSTAR (a version of the Monash stellar evolution code with diffusive mixing; \citealt{Lattanzio86, Frost96, campbellThesis, Doherty14}). \citet{Siess07evolutionII} uses STAREVOL (\citealt{Forestini94low, Siess97synthetic, Siess00internet}) to find $M_{\rm mas}$ for $Z=1\times 10^{-5}$ to $0.04$. \citet{Ibeling13} utilize \texttt{MESA} to calculate $M_{\rm mas}$ for completely metal-free initial compositions to $Z=0.04$. 

Various groups have attempted to constrain $M_{\rm mas}$ with observation. \citet{Smartt09progenitors} use direct imaging of SNe progenitors: they compare observed progenitor luminosities against theoretical models to derive the initial progenitor masses. When using STARS models, they find $M_{\rm mas}=8.5^{+1.0}_{-1.5}$\(M_\odot\). They update their work in \citet{smartt2015observational} using new STARS, Geneva and KEPLER models. They find $M_{\rm mas}=9.5^{+0.5}_{-2}$ with STARS and Geneva, and $M_{\rm mas}=10^{+0.5}_{-1.5}$ with KEPLER. A different observational approach is taken by \citet{diaz2018progenitor} and \citet{diaz2021progenitor}, who date the surrounding populations of SN remnants (former) or historic SN themselves (latter). They quote values of $M_{\rm mas}=7.33^{+0.02}_{-0.16}$\(M_\odot\) and $M_{\rm mas}=8.6^{+0.37}_{-0.41}$\(M_\odot\), respectively. We note that these values are stated with no metallicity dependence, likely given the very few relative numbers of SNe and the complexity of obtaining direct detections.

New and impending observational missions like LSST \citep{LSST}, JWST \citep{gardner2006james}, Gaia \citep{GaiaEDR3}, TESS \citep{TESS}, and Roman \citep{eifler2021cosmology} are generating an extraordinarily rich data climate that, in turn, is driving renewed interest in theoretical stellar structure and evolution across the stellar mass spectrum. In light of recent revolutions in the precision and quantity of observational benchmarks, it is timely and necessary to revisit the physics of stellar interiors and critically examine how we model the processes taking place there. Our theoretical prescriptions underpin not only stellar evolution calculations, but all higher-order models (e.g. of stellar populations, galactic chemical evolution, population synthesis) on which they rely. In order to make the best possible use of the scientific opportunities that will be provided by, e.g., Roman, it is critical that we resolve any tensions regarding key evolutionary parameters---such as $M_{\rm mas}$---in the models we use to interpret observational data.

We attempt to address this in two ways. Firstly, where the metallicity domains overlap, we compare new calculations performed with \texttt{MESA} (version r23.05.1) with current literature. The models we include in our comparison cover a wide range of modelling input physics. We summarize these differences in Table. \ref{physics_comparison} and discuss the impact that variations in these parameters can have on final values for $M_{\rm mas}$. Secondly, we extend the parameter space of $M_{\rm mas}$. There is currently no existing literature that provides $M_{\rm mas}$ for regions with $Z > 0.05$. We calculate $M_{\rm mas}$ for $2.5\times 10^{-3} \geq Z \leq 0.10$. Our model formulation is described in the first paper of this series, \citet{BridgingIcinquegrana}, with additional details on opacities and the treatment of convection provided in \citealt{Cinquegrana22solarcal}. We briefly review these inputs in the next section, \S~\ref{sec:methods}, and discuss any variation between those and our current models. We present our results in \S~\ref{sec:results} and compare them to the available literature in \S~\ref{sec:literaturecomparison}. 

\section{Methods}
\label{sec:methods}

The physics and numerics adopted in the present analysis are largely the same as in \citet{BridgingIcinquegrana} (hereafter, Paper I). Here, we review the settings but refer the reader to Paper I for justification of modeling choices and further details. Some additions and adjustments in our \texttt{MESA} \texttt{inlists} were required given the higher mass regime and correspondingly different physics explored here. Modifications included moving to the most recent, stable release of \texttt{MESA} (version r23.05.1). This allowed us to utilize the new equation of state (EOS) prescription, \texttt{Skye} \citep{jermyn2021skye} (see end of this section for more detail).

Our models are run from the zero-age main sequence (ZAMS) to core C depletion, which we define cross-consistently as the evolutionary point at which the central C mass fraction drops below $X_c = 1\times 10^{-3}$. They are classified as Fe CCSN progenitors if the ONe core is greater than 1.37\(M_\odot\) at core C depletion. This condition is based on the definition of \citet{Nomoto84evolution}, which has since been confirmed in works such as \citet{jones2013advanced}. Otherwise, we consider the models super asymptotic giant branch stars (SAGBs), which typically end their lives as ONe WDs. \footnote{SAGBs can potentially endure a SNe via electron capture onto $^{24}$Mg and $^{20}$Ne \citep{Miyaji80supernova, Nomoto84evolution, Eldridge04prog, jones2013advanced, Jones14, Woosley15remarkable, Doherty17}. For our purposes, we do not consider their case here.} Our core science models cover the metallicity range $Z=0.015$ to $0.10$ in steps of 0.01, corresponding roughly from [Fe/H] = 0.05 to 0.88 when using $Z_{\odot}=0.013$ \citep{Lodders03} or [Fe/H] = -0.04 to 0.78 with the more recent solar abundance, $Z_{\odot}=0.0165$, measured by \citealt{magg2022observational}).

We include some lower metallicity cases, $Z=2.5\times 10^{-3}, 5.0\times 10^{-3}, 7.5\times 10^{-3}$ and $1.5\times 10^{-2}$ ([Fe/H] down to -0.75 or -0.82, depending on $Z_{\odot}$), for the purpose of comparison to other works. Our initial grid covers 9 to 10\(M_\odot\) in steps of 0.1$M_{\odot}$, increasing in resolution to steps of 0.05$M_{\odot}$ as the border between intermediate and massive fates becomes apparent. 

We calculate an initial helium abundance based on the initial metallicity of the models as follows: 
\begin{equation} \label{delYdelZ}
    \rm{Y}_i = Y_0 + \frac{\Delta Y}{\Delta Z} \times \rm{Z}_i.
\end{equation}
Here, $Y_0$ is the primordial He abundance, which has an observed value of $Y_0=0.2485$ \citep{Aver13}. With time, the helium abundance increases at the rate of the He-to-metal enrichment ratio, $\frac{\Delta Y}{\Delta Z}=2.1$ \citep{Casagrande07}. The resultant He mass fractions are listed in Table \ref{initialquantities}.

\begin{table*}
\centering
\caption{Initial composition of our 1D stellar evolution models. [Fe/H] is often used as a proxy for metallicity in observational data, we provide a rough conversion based on two different solar abundances with $\rm [Fe/H] \sim \log_{10}(\frac{Z}{Z_{\odot}})$.}
\label{initialquantities}
\begin{tabular}{|c|c|c|c|c|c|} \hline \hline
X (Hydrogen) & Y (Helium) & Z (Metals) & Z (Metals)& [Fe/H] & [Fe/H] \\ 
 &  &  &  & $Z_{\odot}=0.0133$  & $Z_{\odot}=0.0165$ \\ 
  &  & &  & \citep{Lodders03} & \citep{magg2022observational} \\ \hline
0.744 & 0.254 & 0.0025 & 0.25\%    &   -0.73    &-0.82\\
0.736 & 0.259 & 0.005 & 0.5\%      &   -0.42	    &-0.52\\
0.729 & 0.264 & 0.0075 & 0.75\%    &   -0.25    &-0.34\\
0.705 & 0.280 & 0.015 &1.5\%       &    0.05	   &-0.04\\
0.689 & 0.291 & 0.02 &2\%          &   0.18    &0.08\\
0.658 & 0.312 & 0.03 &3\%          &   0.35	    &0.26\\
0.627 & 0.333 & 0.04 &4\%          &   0.48    &0.38\\
0.596 & 0.354 & 0.05 &5\%          &   0.58	    &0.48\\
0.565 & 0.375 & 0.06 &6\%          &   0.65	    &0.56\\
0.534 & 0.396 & 0.07 &7\%          &   0.72	    &0.63\\
0.503 & 0.417 & 0.08 &8\%          &   0.78	    &0.69\\
0.472 & 0.438 & 0.09 &9\%          &   0.83    &0.74\\
0.441 & 0.459 & 0.10 &10\%         &   0.88    &0.78\\
\hline \hline 
\end{tabular}
\noindent 
\end{table*}

We use the \verb|approx21_cr60_plus_co56| nuclear reaction network. APPROX21 is a modified version of the original APPROX19 network \citep[see][]{Weaver78presupernova}. This contains the isotopes required for H burning to Si burning: $^{1}$H, $^{3}$He, $^{4}$He, $^{12}$C, $^{14}$N, $^{16}$O, $^{20}$Ne, $^{24}$Mg, $^{28}$Si, $^{32}$S, $^{36}$Ar, $^{40}$Ca, $^{44}$Ti, $^{48}$Cr, $^{52}$Fe, $^{54}$Fe, $^{56}$Ni, neutrons and protons, with the addition of $^{56}$Cr and $^{56}$Fe. Finally, \verb|approx21_cr60_plus_co56| utilises the isotopes listed above, but also follows $^{60}$Cr and $^{56}$Co\footnote{Further discussion of the alpha-chain reaction networks in \texttt{MESA} can be found at \href{https://cococubed.com/code_pages/burn_helium.shtml}{cococubed}}. The reaction rates we use in our \texttt{MESA} calculations are sourced from JINA REACLIB \citep{Cyburt10}, with some additional weak rates from \citet{Fuller85stellar, Oda94rate, Langanke00shell}. 

We use the Mixing Length Theory (MLT; \citealt{Prandtl25, Bohm1958, Boehm-Vitense79, Paxton10instrument1}) of convection to model energy transport in superadiabatic regions. We use the \citet{Henyey65mlt} prescription, which is a slightly modified version of \citet{Vitense53}. In \citet{Cinquegrana22solarcal}, we calibrated the MLT parameter, $\alpha_{\rm MLT}$, to be $\alpha_{\rm MLT}=1.931$ for our choice of input physics (for a thorough review of this topic, see \citealt{JoyceTayar2023}). We determine convective stability with the Ledoux criterion and define the borders between convective and radiative regions using the predictive mixing algorithm \citep{Paxton18instrument4, Constantino15tre, Bossini15unc}. 

We use OPAL opacities \citep{Iglesias96} for high temperature regions and custom {\AE}SOPUS tables \citep{Marigo09} for low temperature regions. Details on these tables can be found in \citet{Cinquegrana22solarcal}, with specific parameters listed in the appendix of the arXiv\footnote{\href{https://arxiv.org/pdf/2204.08598.pdf}{https://arxiv.org/pdf/2204.08598.pdf}} version. We model the atmospheric boundary conditions using a gray $t$--$\tau$ relation with Eddington integration. 

Mass loss on the red giant branch is modelled using the \citet{Reimers75} approximation, with $\eta _{\rm reimers}=0.477$ \citep{Mcdonald15mass}. We use \citet{Blocker95a} for mass loss along the asymptotic giant branch (AGB), with $\eta _{\text{Bl\"{o}cker}}=0.01$. 


In Paper I, we use the \texttt{MESA} equation of state (EOS), which is a blend of OPAL \citep{Rogers2002}, SCVH \citep{Saumon1995}, FreeEOS \citep{Irwin2004}, HELM \citep{Timmes2000}, and PC \citep{Potekhin2010}. With the release of r22.05.1 came a new EOS prescription, \texttt{Skye} \citep{jermyn2021skye}, which models fully ionized matter (see \texttt{MESA} VI; \citealt{Jermyn22instrument6}). This change was adopted in part because we encountered convergence issues in our models for higher initial masses due to blending between the HELM and PC prescriptions. Utilising \texttt{Skye} instead has removed these issues. 

Another difference between the modeling configuration adopted in Paper I and the present study is use of the Ledoux convective stability criterion rather than Schwarzschild. 
Using Ledoux allows us to model semiconvection \citep{schwarzschild1958evolution} and thermohaline mixing \citep{Stern60thermohaline} during the core He burning phase--processes which become more important in this mass regime \citep{Thomas67thermohaline, ulrich1972thermohaline}. 
We use the \citet{Langer85evo} prescription for semiconvection and \citet{kippenhahn1980time} for thermohaline convection. Their efficiency parameters, \verb|alpha_semiconvection| and \verb|thermohaline_coeff|, are set to 0.01 and 2, respectively, for the evolution up to the end of core He burning (see discussion in \citealt{TayarJoyce22} for \verb|thermohaline_coeff|). 

We use a diffusive overshooting scheme based on \citet{Herwig00}\footnote{The \texttt{MESA} implementation of which is described in \citet{Paxton10instrument1}}. During the evolution from ZAMS to terminal age core He burning (TACHeB), we adopt a phase--specific value of \verb|overshoot_f|$=0.01$ and \verb|overshoot_f0|$=0.005$ for convection zones of all types, 
as used in \citet{farmer2019mind, marchant2019pulsational, renzo2020predictions, renzo2020sensitivity}. For evolutionary phases that do not result in significant core growth (pre main sequence, post TACHeB), these values are reduced to \verb|overshoot_f = 0.005d0| and \verb|overshoot_f0 = 0.001d0|, respectively, to aid with convergence. These numbers are similar to those used in \citet{farmer2016variations}, whose choices were driven by the 1D to 3D calibration performed by \citet{jones2016idealised}. 

We note that the timestep and solver controls in our inlists are based on the template \texttt{test\_suite} case \verb|12M_pre_ms_to_core_collapse| by R. Farmer, available in \texttt{MESA} version r22.05.1. These settings were calibrated by \citet{farmer2016variations} and \citet{laplace2021different}\footnote{Repositories for those papers can be found at 10.5281/zenodo.2641723 and 10.5281/zenodo.5556959}. 

\section{Results} 
\label{sec:results}

\begin{figure*} 
    \centering
	\includegraphics[width=16cm]{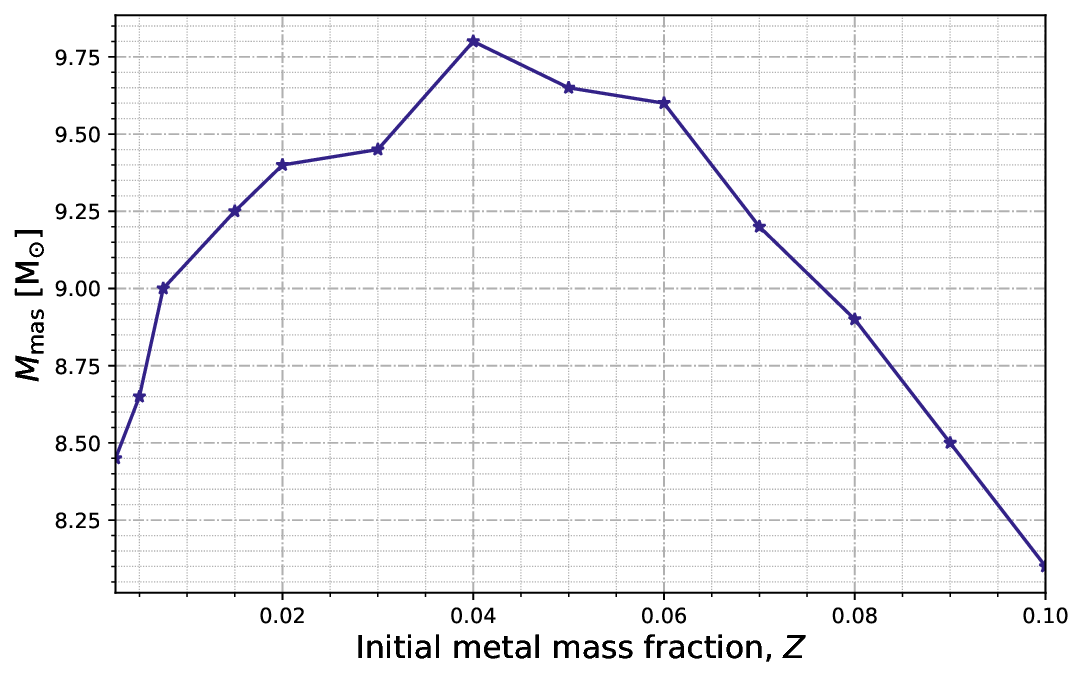}
    \caption{Initial mass to produce a Fe CCSNe, $M_{\rm mas}$, as a function of initial metallicity. These models were calculated with \texttt{MESA} version r23.05.1 and classified as massive if they have ONe cores greater than 1.37\(M_\odot\) \citep{Nomoto84evolution}. They have canonical helium values calculated with the $\frac{\Delta Y}{\Delta Z}$ law. 
    We model convection with the MLT, where $\alpha_{\rm MLT}=1.931$ (calibrated in \citealt{Cinquegrana22solarcal}) and use OPAL opacities for high temperature regions, custom {\AE}SOPUS tables for low temperature regions. We model mass on the red giant branch using \citet{Reimers75} and \citet{Blocker95a} for mass loss along the asymptotic giant branch (AGB). We use the Ledoux criteria for convective stability, allowing us to include semiconvection and thermohaline mixing. Finally, we use the predictive mixing algorithm to find the borders between convective and radiative regions, and include convective overshooting using the prescription of \citet{Herwig00}.}
    \label{fig:ourresults}
\end{figure*}

\begin{table*}
\centering
\caption{$M_{\rm mas}$ values in [\(M_\odot\)] as a function of initial metal mass fraction, $Z$.}
\label{tab:results}
\begin{tabular}{|c|c|c|c|c|c|c|c|c|c|c|c|c|c|} \hline \hline
$Z \rightarrow$ & 0.0025 & 0.005 & 0.0075 & 0.014  & 0.02 & 0.03 & 0.04 & 0.05 & 0.06 & 0.07 & 0.08 & 0.09& 0.10 \\ 
 References   &    &   &   & to &   &   &   &   &   &  &    &  &   \\ 
  $\downarrow$  &    &   &   & 0.015  &   &   &   &   &   &  &    &  &   \\ \hline
  This work  &  8.45  & 8.65  & 9.0  & 9.25  &  9.4 &  9.5 &  9.8 &  9.65 &  9.6 & 9.2  & 8.9   & 8.5 & 8.1 \\ 
  \citet{Eldridge04prog}, no OS  &  --  &  -- &  -- &  -- & 9.5   & 10  & 10  &   10 &  -- & -- &   -- & -- &  -- \\ 
   with OS &  --  & --  & --  &  -- &  7 &  7.5 & 7.5  &  7.5 & -- & -- &  --  &  --&  -- \\ 
  \citet{Siess07evolutionII}, no OS  &  --  & --  &  -- & --  & 10.93  &  -- &  10.89 & --  &  -- &  --& --   & -- & --  \\ 
   with OS &  --  &  -- & --  &  -- & 8.83  & --  &  -- & --  &   --& -- & --   & -- &  -- \\ 
  \citet{Poelarends08supernova}, STERN  &  --  &  -- &  -- & --  &  13 &  -- & --  &  -- &--   & -- & --   & -- &  -- \\ 
   EVOL &   -- &  -- &  -- & --  &  10.5 &  -- & --  &  -- &  -- &  --&  --  & -- & --  \\ 
   KEPLER &   -- & --  &  -- & --  &  9.2 & --  & --  &  -- & --  &  --&  --  & -- & --  \\ 
  \citet{Ibeling13}  &  8.65  &  8.95 & 9.05  &  9.35 & 9.5  &  9.6 &  9.85 &  -- &  -- & -- &  --  & -- &  -- \\ 
  \citet{jones2013advanced}  &  --  &  --& --  & 9.5  & --  &  -- & --  &  -- &  -- & -- &  --  & -- &  -- \\  
  \citet{Doherty15super}  &  --  &  -- & --  & --  &  9.9 &  -- &  -- &  -- & --  &  --& --   & -- & --  \\ 
  \citet{Woosley15remarkable}  & --   & --  &  -- & 9  &  -- & --  & --  &  -- & --  & -- & --   & -- & --  \\ 
\hline \hline 
\end{tabular}
\noindent 
\end{table*}

\begin{table*}
\centering
\caption{Relevant input physics for the models shown in Figure. \ref{fig:comp}. Here, CT = treatment of convection,  CS = convective stability criterion, CBA = convective boundary placement algorithm, SC = semiconvection, NG = data not provided.}
\label{physics_comparison}
\begin{tabular}{|c|c|c|c|c|c|c|c|} \hline \hline
Reference & Symbol  & Code & Z$_{\rm max}$ & CT & CS & CBA & SC \\ \hline
This work & \includegraphics[scale=0.25]{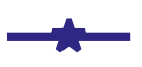} & \texttt{MESA} & 0.10 & MLT & Ledoux & \citet{Herwig00} & Yes \\
 &  &  &  & $\alpha_{\rm MLT}$=1.931 &  & & \\
\citet{Eldridge04prog}& \includegraphics[scale=0.25]{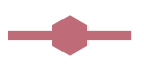} & STARS & 0.05 & MLT & (Modified) schwarzschild$^{\star}$ & \citet{schroder1997critical} & Yes \\
 &  &  &  & $\alpha_{\rm MLT}$=2.0 &  &  & \\
--- & \includegraphics[scale=0.25]{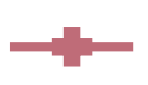} & --- & --- & --- & --- & No & --- \\
\citet{Siess07evolutionII} & \includegraphics[scale=0.25]{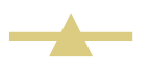} & STAREVOL & 0.04 & MLT & Schwarzschild & No & No \\
 &  &  &  & $\alpha_{\rm MLT}$=1.75 &  &  & \\
---& \includegraphics[scale=0.25]{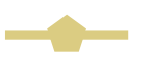} & --- & --- & --- & --- & \citet{freytag1996hydrodynamical} & --- \\
 &  &  &  &  &  & \citet{blocker1998lithium} & \\
\citet{Poelarends08supernova} & \includegraphics[scale=0.25]{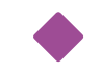} & STERN & 0.02 & MLT & Ledoux & NG & Yes \\
---& \includegraphics[scale=0.25]{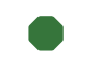} & EVOL & 0.02 & MLT & Schwarzschild & \citet{Herwig00} & No \\
--- & \includegraphics[scale=0.25]{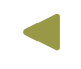} & KEPLER & 0.02 & MLT & Ledoux & Yes & Yes \\
\citet{Ibeling13} & \includegraphics[scale=0.25]{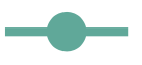} & \texttt{MESA} & 0.04 & MLT & NG & \citet{freytag1996hydrodynamical} & NG \\
 &  &  &  &  &  & \citet{Herwig00} & \\
\citet{jones2013advanced} & \includegraphics[scale=0.15]{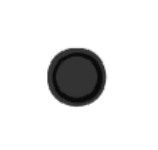} & \texttt{MESA} & 0.014 & MLT & Schwarzschild$^{\dagger}$ & \citet{freytag1996hydrodynamical} & No \\
 &  &  &  & $\alpha_{\rm MLT}$=1.73 &  & \citet{Herwig00} & \\
 
\citet{Doherty15super} & \includegraphics[scale=0.25]{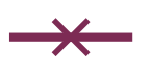} & MONSTAR & 0.02 & MLT & Ledoux & \citet{Lattanzio86} &  NG\\
 &  &  &  & $\alpha_{\rm MLT}$=1.75 &  &  & \\
\citet{Woosley15remarkable} & \includegraphics[scale=0.08]{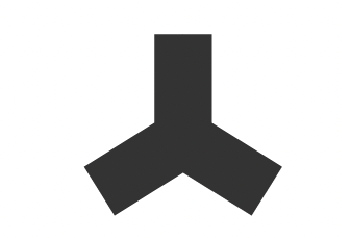} & KEPLER & 0.015 & MLT & Ledoux & Yes &  Yes\\ 
\hline \hline
\end{tabular}
\medskip\\
\noindent $\star$ \citet{Eldridge04prog} use a modified version of the Schwarschild criterion with an extra term that allows for the modelling of semi-convection and diffusive overshooting. See \S~ 2.1.2 of \citet{Eldridge04prog} and \citet{schroder1997critical}. \\
\noindent $\dagger$ \citet{jones2013advanced} use the Schwarschild criterion for the majority of their evolution, but switch to Ledoux during late stage evolution for some models.
\end{table*}

\begin{figure*} 
	\includegraphics[width=18cm]{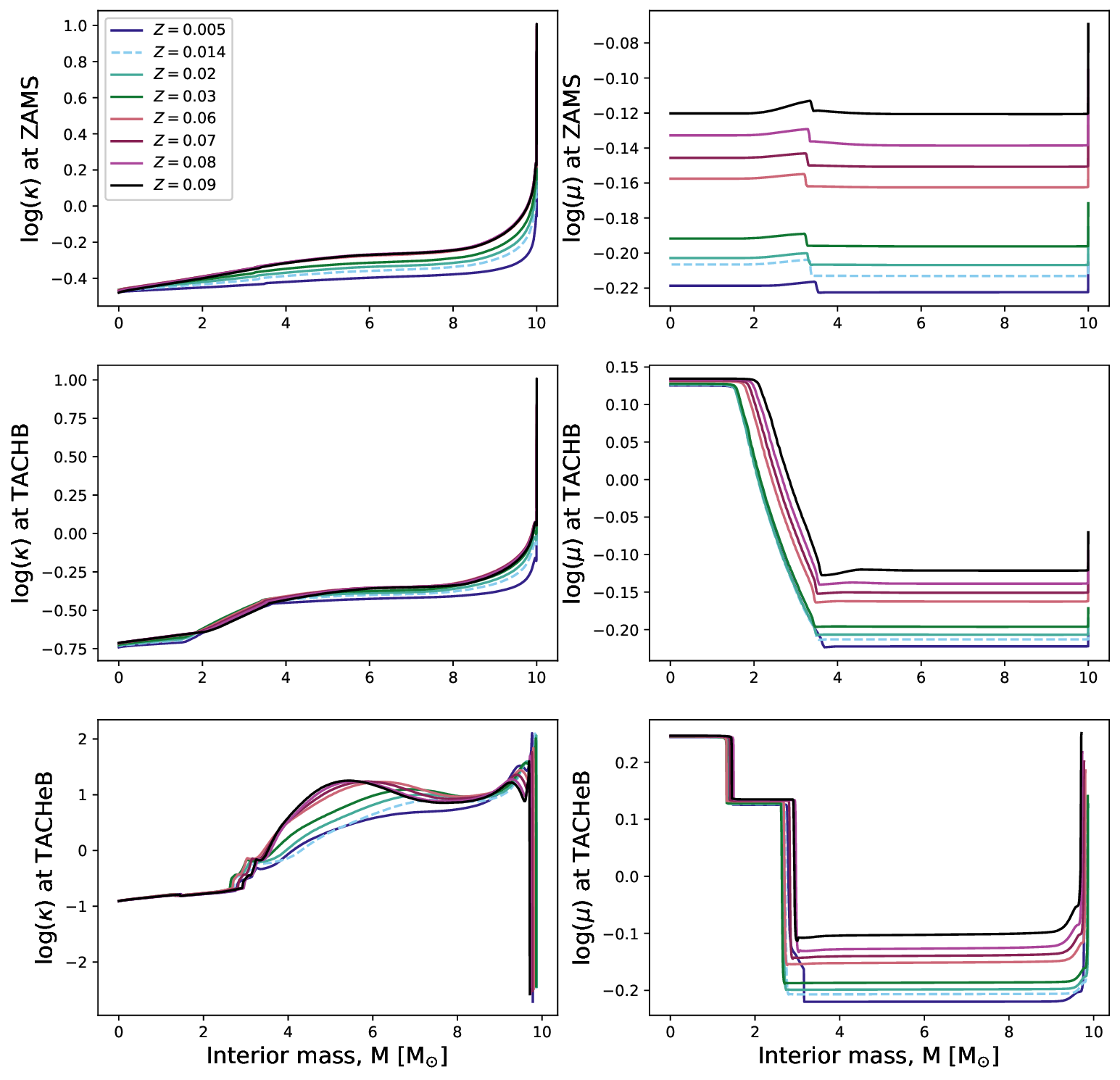}
    \caption{The $\kappa$ and $\mu$ profiles of a set of 10\(M_\odot\) models. We vary $Z$ between $Z=0.005$ and $Z=0.09$. Snapshots are provided at the zero-age main sequence, terminal age core H burning and terminal age core He burning (when $^{1}$H and $^{4}$He drop below $1 \times 10^{-3}$).}
    \label{fig:mukappa}
\end{figure*}

\begin{figure} 
	\includegraphics[width=9cm]{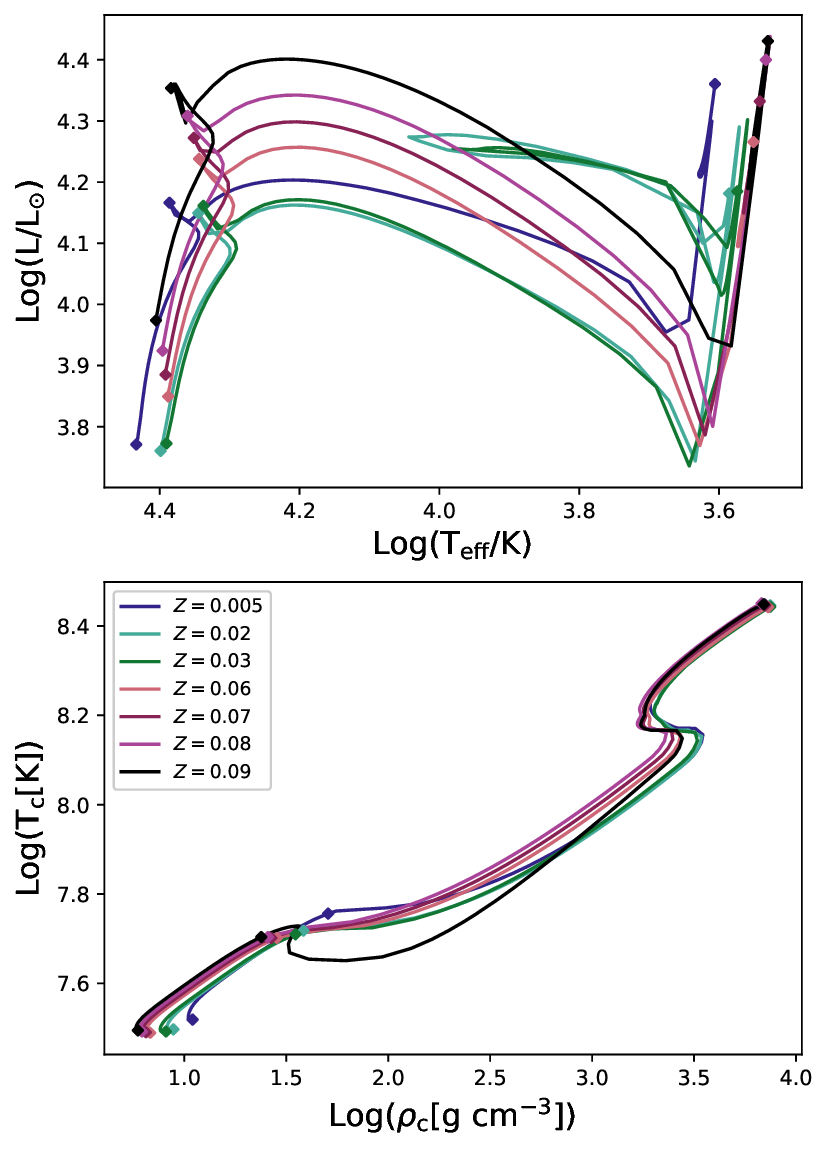}
    \caption{The stellar tracks (upper panel) of 10\(M_\odot\) models, $Z$ between $Z=0.005$ and $Z=0.09$, and their $T_c$ and $\rho_c$ behaviour (lower panel).}
    \label{fig:4mgrid}
\end{figure}

Our primary results are presented in Figure. \ref{fig:ourresults} and Table. \ref{tab:results}. We find higher initial masses are required for models to undergo Fe CCSNe as the initial metallicity increases from $Z \approx 1\times 10^{-3}$ to $\approx2$\(Z_\odot\). That is, the magnitude of the H exhausted core mass for models in this range decreases with increasing $Z$. For $Z>0.05$ (which also corresponds to the maximum metallicity currently tested in the literature), these trends {reverse}. With increasing metallicity, the H exhausted core mass now begins to increase again, so \textit{lower} initial masses will undergo Fe CCSNe as metallicity increases to $Z=0.10$. 

It is important to understand why this trend reversal occurs at the highest metallicities we consider. As the initial metal content in a gas increases, so too do the opacity ($\kappa$) and the mean molecular weight ($\mu$) of the gas. If we consider the mass luminosity relation, 
\begin{equation} \label{masslumrelation}
    L \propto \frac{\mu ^4 M^3}{\kappa}, 
\end{equation}
and the Stefan-Boltzmann relationship, 
\begin{equation} \label{stefanboltzman}
    L = 4 \pi R^2 \sigma T_{\rm eff}^4, 
\end{equation}
we can see that an increasing $\kappa$ leads to lower luminosities and effective temperatures. Yet, an increasing $\mu$ leads to higher luminosities and effective temperatures. Although Equations \ref{masslumrelation} and \ref{stefanboltzman} are simplifications of the treatments in \texttt{MESA} and other stellar evolution codes, the balance between $\kappa$ and chemical composition remains relevant, and we see these opposing effects of $\kappa$ vs $\mu$ dominate for different metallicity ranges. 

The impact of the higher $\kappa$ and $\mu$ in our metal-rich models manifests in a variety of stellar processes. For the following discussion, we consider the metal-rich range to extend from $Z=0.014$ (roughly solar metallicity) to $Z=0.10$ (10\% metals). In Figure \ref{fig:mukappa}, we show the $\kappa$ and $\mu$ profiles of a set of 10\(M_\odot\) models, we vary $Z$ between $Z=0.005$ and $Z=0.09$. Snapshots are provided at the ZAMS, terminal age core H burning (TACHB) and TACHeB. 

At the ZAMS, $\kappa$ and $\mu$ in the stellar envelope increases with initial metallicity. The opacity reaches a plateau in the higher metallicity range, with the $Z=0.06, 0.07, 0.08$ and $Z=0.09$ models achieving the same values. At the end of core H burning, the variation in $\kappa$ (between the models) in the envelope has decreased, but there still remains a significant difference between the $\mu$ curves in the stellar envelopes of the different models. It appears that an increasing gas metal content primarily impacts the envelopes of these models. The feature in the $\kappa$ plot at the TACHeB is the Fe $\kappa$ bump, identified in the early 1990s \citep{Rogers92, seatonopacity}, which occurs at $\log T\approx 5.3$K. 
%

In Figure. \ref{fig:4mgrid}, we show the stellar tracks of these same models as well as their central temperatures ($T_c$) and densities ($\rho_c$). The $Z=0.005, 0.02$ and $0.03$ models evolve onto ZAMS with roughly the same luminosity, but the $Z=0.02, 0.03$ cases emerge with lower effective temperatures. These two models maintain lower $L$ and $T_{\rm eff}$, in comparison to the $Z=0.005$ model, for their evolution up to TACHeB. They begin their ascent of the red giant branch at lower $L$ ($\Delta L \approx 0.25$\(L_\odot\)) and experience much longer blue loops (see \citealt{walmswell2015blue}) than any of the other models. These loops are very sensitive to microphysics choices (e.g. the $^{14}$N($p, \gamma)^{15}$O rate, see \citealt{weiss2005influence}). Consequently, there remains a lack of consensus in the literature with respect to how the loops \textit{should} behave as their initial metal content is varied. The models with $Z \geq 0.06$ initiate (and complete) core H burning (and He burning) at $L$ values that increase with $Z$. In terms of their central conditions, all the metal-rich models show slightly cooler temperatures and lower densities than the $Z=0.005$ case at ZAMS and TACHB but evolve \textit{through} the phases with slightly warmer tracks. All models complete core He burning at approximately the same temperature and density. 

To summarize, as metallicity increases, both the opacity and mean molecular weight of the gas also increase. The difference in $\kappa$ and $\mu$ between the $Z=0.005$ case and the metal-rich models is most significant in the stellar envelope. At the lower end of the metal-rich range ($Z=0.02, 0.03$), the models experience cooler and slightly less luminous core H and He burning phases, which leads to H exhausted core masses that decrease in magnitude with $Z$. Consequently, these models require higher initial masses for key burning phases, to make up for their less massive cores. The models at the higher end of the metal-rich range ($Z\geq0.06$) experience more luminous and warmer main sequence burning lifetimes, but still have cooler and less luminous surfaces as they ascend the giant branches. The conflicting behaviour of these metal-rich models is driven by the competition between the effects of the higher $\kappa$ and $\mu$. The most enriched $Z$ models possess greater initial He levels, represented by the $\frac{\Delta Y}{\Delta Z}$ law (see Equation. \ref{delYdelZ}). Higher $Y_i$ further drives up $\mu$, and the associated properties with a higher $\mu$. Namely, the higher $T_{\rm eff}$ and $L$ lead to more efficient core burning phases and thus more massive H exhausted cores. Consequently, these models require lower initial masses for higher order burning stages with increasing $Z$ (e.g. core C ignition). 

Low and intermediate mass metal-rich models experience mass loss rates that increase with $Z$. \citet{Karakas22paper1} and \citet{Cinquegrana22paper2}, found the same models show drastically reduced mixing efficiencies for third dredge up (TDU) episodes on the thermally pulsing AGB (TP-AGB). Although low mass models will still endure the first dredge up -- and intermediate models the second dredge up as well --- the first two dredge up events only mix H burning products up to the stellar surface. The TDU is the first opportunity for these models to mix He burning products and potentially elements produced by the slow neutron capture process (\textit{s}-process) up to the surface. Thus, eventually those products can be contributed back to the interstellar medium as the envelope is eroded by stellar winds, rather than remaining locked inside the white dwarf remnant. Due to their lower effective temperatures, very metal-rich models are also often unable to ignite H at the base of their convective envelopes on the TP-AGB, known as hot bottom burning. The higher mass loss rates of the models shortens their AGB lifetimes and thus they have less opportunities to experience their less efficient TDU and hot bottom burning episodes. Interestingly, we found that the \textit{second} dredge up episode (for intermediate mass stars, on the early red giant branch) is largely metallicity independent (see \citealt{Karakas22paper1}). Thus even though the net yields of these models are depleted in He burning and \textit{s}-process products, they are \textit{rich} in secondary H burning nucleosynthesis products. 


We end this discussion by emphasizing that all the downstream implications of a high $Z$ (e.g. faster mass loss rates, lower TDU efficiencies) increase monotonically with $Z$, regardless of the competition between $\kappa$ and $\mu$. The core mass, however, does not.

Lastly, it is worth noting that the reversal in trend we see for $M_\text{mas}$ is sensitive to the scaling of the initial He abundance with the initial $Z$ abundance. Whether He does scale linearly with metallicity (and by the same rate in all regions of the universe) is unknown, as is the exact value of the primordial He abundance, $Y_0$. Perhaps most uncertain is the value of $\frac{\Delta Y}{\Delta Z}$. In this work, we use $\frac{\Delta Y}{\Delta Z}=2.1$ \citep{Casagrande07}, but this quantity is very difficult to constrain. To do so, we need to measure the stellar He abundance directly, which is only feasible in stars with $\rm T_{eff} > 8000$K (i.e. main sequence stars with initial masses greater than 1.5\(M_\odot\); \citealt{valcarce2013fundamental}). A range of methods to estimate $\frac{\Delta Y}{\Delta Z}$, both observational and theoretical, yield values of $\frac{\Delta Y}{\Delta Z}$ between 0.7 to 10 \citep[e.g.][]{faulkner1967quasi, perrin1977fine, renzini1994hint, fernandes1996width, pagel1998delta, ribas2000chemical, chiappini2002evolution, jimenez2003cosmic, salaris2004initial, balser2006chemical, izotov2007primordial, casagrande2007helium, gennaro2010deltay, portinari2010revisiting, brogaard2012age, martig2015young, Joyce18class}. This uncertainty aside, the precise value of $\frac{\Delta Y}{\Delta Z}$ will have no effect on the qualitative trend. The only way this could have an impact would be if this value were not the same universally.


\section{Literature comparison}\label{sec:literaturecomparison}

\begin{figure*} 
    \centering
	\includegraphics[width=18cm]{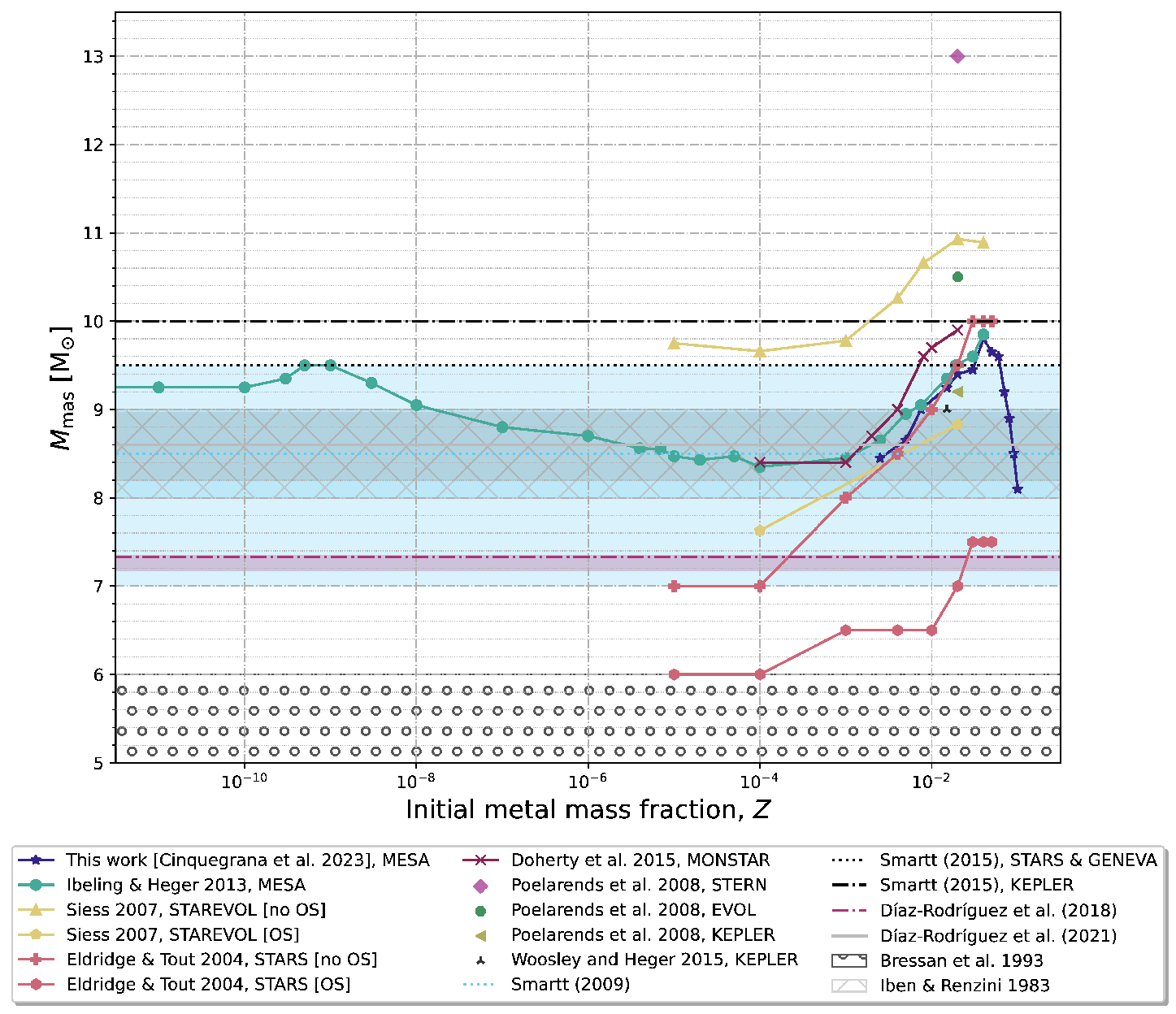}
    \caption{A comparison of our calculated $M_{\rm mas}$ against that currently available in the literature, where $M_{\rm mas}$ is the minimum initial mass required for a star to undergo an Fe CCSNe. A copy of this figure using a linear scale, without shading is included in the appendix, \ref{fig:append_comp}. The relevant input physics used in each set of models, where available, is listed in Table. \ref{physics_comparison}. Isolated curves are theoretically derived $M_{\rm mas}$ values. The hatched regions (corresponding to \citealt{iben1983asymptotic} and \citealt{bressan1993evolutionary} are metallicity independent approximations for $M_{\rm mas}$, often used in galactic chemical evolution simulations. \citet{Smartt09progenitors}, \citet{diaz2018progenitor} and \citet{diaz2021progenitor} are $M_{\rm mas}$ values derived using observational quantities. The upper and lower uncertainties for these measured quantities are indicated by the shaded regions in the same colour.}
    \label{fig:comp}
\end{figure*}

In Figure \ref{fig:comp}, we compare our results to the theoretically derived results of \citet{Eldridge04prog}, \citet{Siess07evolutionII}, \citet{Poelarends08supernova}, \citet{Ibeling13}, \citet{jones2013advanced}, \citet{Doherty15super} and \citet{Woosley15remarkable}. We also highlight the mass regions that are commonly used in galactic chemical evolution simulations (\citealt{iben1983asymptotic, bressan1993evolutionary} are theoretically derived; \citealt{Smartt09progenitors} is derived using observational quantities). A version of this figure using a linear scale and without shading is included in the appendix, \ref{fig:append_comp}.
Our values match those of \citet{Ibeling13} very well across the entire metallicity range. There is a maximum variation of 3.4\% between our $M_{\rm mas}$ values; on average they differ by $\sim 1.4$\%. This difference extends to 1-5.2\% between our results and those of \citet{Eldridge04prog} models \textit{with no} overshoot, \citet{Woosley15remarkable}, \citet{jones2013advanced}, \citet{Doherty15super} and the \citet{Poelarends08supernova} KEPLER models. There is a 6.3\% deviation between ours and \citet{Siess07evolutionII} \textit{with} overshoot.   
The \citet{Siess07evolutionII} models \textit{with no} overshoot and the EVOL models from \citet{Poelarends08supernova} calculate significantly higher values for $M_{\rm mas}$. There is a 10-15\% difference between those models and ours.
\citet{Eldridge04prog} models \textit{with} overshoot and the \citet{Poelarends08supernova} STERN model show the largest variation, approximately 23-32\% between those and our values for $M_{\rm mas}$. 
Regardless of the quantitative difference in values between the models, we note that the overall shapes of the $M_{\rm mas}-Z$ curves are very similar. 

We also compare our models and calculations in the literature against the shaded regions, representing common non-metallicity-dependent approximations for $M_{\rm mas}$ used in galactic chemical evolution simulations. The \citet{iben1983asymptotic} range ($8-9$\(M_\odot\)) encompasses the majority of the curves for $1\times 10^{-3} < Z < 0.01$. It also contains our highest metallicity models, $Z=0.09$ and $0.10$. The \citet{Smartt09progenitors} range is larger, 8$\pm$1\(M_\odot\), and also contains the \citet{Eldridge04prog} models with no overshoot for $1\times 10^{-5} < Z < 1\times 10^{-3}$. The \citet{bressan1993evolutionary} range ($5-6$\(M_\odot\)) misses all but the \citet{Eldridge04prog} overshoot values for $Z=1\times 10^{-4}$ and $1\times 10^{-5}$. All three completely miss the metallicity range between $Z=1\times 10^{-2.5}$ and $Z=0.08$, which is particularly important for simulations of spiral galaxies with solar (or super solar) metallicity bulges.
Given a typical IMF, such as \citep{Salpeter55}, favour low-mass stars, the consequences of using one of these non-metallicity-dependent ranges is the over prediction of SNe events at solar metallicity and miscalculation of appropriate yields. Likewise, if one uses a metallicity-dependent curve, e.g. the \citet{Ibeling13} values, and extrapolates for $Z>0.04$, then one is likely under-predicting the number of SNe occurring in the most metal-rich regions. 

\section{Key modeling uncertainties}
\label{modelling}
In Table \ref{physics_comparison}, we summarize the various physics prescriptions adopted by each set of authors whose results are shown in Figure \ref{fig:comp}. The physical components described include convective treatment, convective boundary placement algorithms, semiconvection treatment, as well as the evolution code used. We discuss the most critical physical choices and their respective contributions to the uncertainty of $M_{\rm mas}$. We focus on our current understanding of these processes and comment on the validity of their implementation in stellar evolution codes. There are, of course, many other uncertainties related to stellar modelling that we will not discuss here (c.f. \citealt{Choi2018, Tayar2022, Joyce2023}); we have chosen in this study to focus on the assumptions that are most pertinent to our derivation of $M_{\rm mas}$. 

\subsection{Convective boundaries}\label{CB}

The border that divides convective and radiative regions in a stellar interior can be found using the Schwarzschild criterion for dynamic stability, 
\begin{equation}\label{schw}
    \nabla_{\rm rad} < \nabla_{\rm ad}. 
\end{equation}
$\nabla_{\rm rad}$ and $\nabla_{\rm ad}$ represent the radiative and adiabatic temperature gradients. The general temperature gradient, $\nabla$, is defined as:
\begin{equation}\label{generaltemp}
    \nabla = \frac{d \ln T}{d \ln P} = \frac{P}{T}\frac{dT}{dP}.
\end{equation}
One may also use the Ledoux criterion, which reduces to Schwarzschild where there is no composition gradient: 
\begin{equation}\label{ledoux}
    \nabla_{\rm rad} < \nabla_{\rm ad} + \frac{\phi}{\delta} \nabla_{\mu}.
\end{equation}
Here, $\phi$ ($\delta$) represents the density gradient with respect to composition (temperature) and $\nabla_{\mu}$ is the composition gradient. Given that $\nabla_{\rm rad} < \nabla_{\rm ad}$ denotes a region in dynamic stability, $\nabla_{\rm rad} = \nabla_{\rm ad}$ indicates the location at which the convective acceleration reduces to zero. With a proper implementation of the Ledoux and Schwarzschild criteria, both Equations. \ref{schw} and \ref{ledoux} should identify the same border location \citep{gabriel2014proper}. \citet{Anders22schwarzschild} demonstrate this by considering how convective zones interact with Ledoux--stable regions in 3D hydrodynamical simulations. They find that where the evolution timescale is significantly greater than the convective mixing timescale, the convective borders defined by both criteria are equivalent. 

A common approach to locate the convective--radiative boundary in stellar structure calculations 
is to use a sign-change algorithm, as discussed in \citet{gabriel2014proper, Paxton18instrument4} and \citet{Paxton19instrument5}. Let the discriminant $y$ for the Schwarzschild and Ledoux criteria be given by 
\begin{equation}\label{discrim_schw}
    y_s \equiv \nabla_{\rm rad} - \nabla_{\rm ad}, 
\end{equation}
and
\begin{equation}\label{discrim_led}
    y_l \equiv \nabla_{\rm rad} - (\nabla_{\rm ad} + \frac{\phi}{\delta} \nabla_{\mu}), 
\end{equation}
respectively.
%
%
If we were to scan along the radius of a stellar model, the sign of the applicable dynamical stability criterion (either Equation \ref{schw} or \ref{ledoux}) would reverse at the transition from a convective to radiative region. At this point, the relevant discriminant should reduce to zero on both sides of the border. In practice, this radial search is performed first from the convective side of the border, given that the location at which convective acceleration reduces to zero is only meaningful in a convective region.
\begin{figure} 
	\includegraphics[width=8cm]{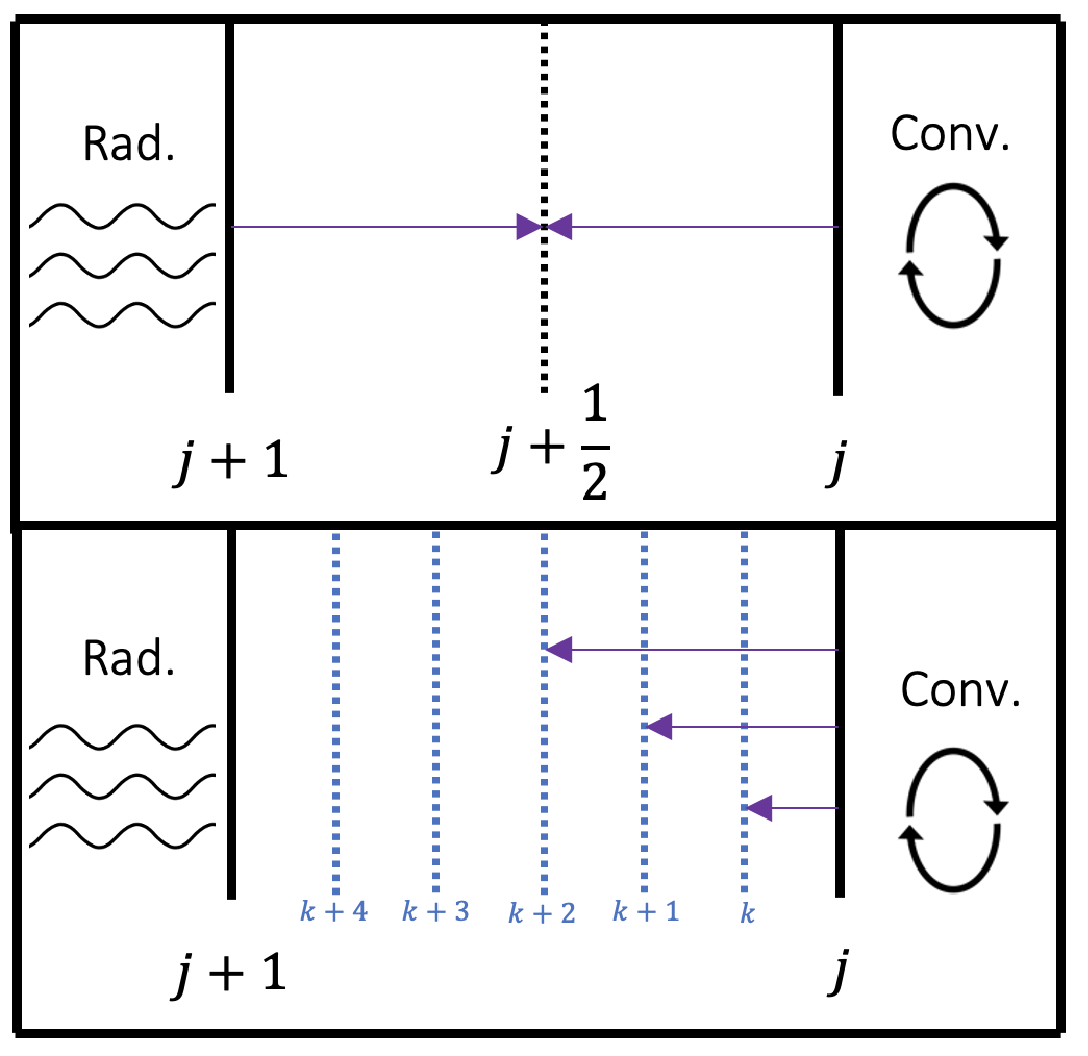}
    \caption{Depiction of the sign-change algorithm (upper panel) and the subgrid physics model (or ``direct search’’ algorithm; lower panel) used in 1D stellar evolution codes to locate the boundaries between convective and radiative regions. $j$ and $j+1$ are the mesh shells which represent the points at which the discriminants ($y_s$ and $y_l$, see Equations. \ref{discrim_schw} and \ref{discrim_led}) first reduce to zero on the radiative and convective sides of the boundary, respectively. $k$ to $k+5$ represent a finer resolution of the region between $j$ and $j+1$.}
    \label{conv_regions}
\end{figure}

We paint a simplified picture of the sign-change algorithm in the upper panel of Figure \ref{conv_regions}. Here, we have located the mesh shells, $j$ and $j+1$. These represent the points at which the discriminant first reduces to zero on the radiative and convective sides of the boundary, respectively. Mesh shell $j$ indicates the location at which the acceleration of the convective eddies reduces to zero, but they will still have some non-zero velocity. Consequently, these eddies may potentially \textit{overshoot} some distance beyond $\nabla_{\rm rad} = \nabla_{\rm ad}$. So, the actual convective border ($v=0$) is commonly adjusted \textit{ad hoc} by interpolating between the two mass shells $j$ and $j+1$. As described in \citet{gabriel2014proper}, this approach is sufficient while using the Schwarzschild criterion with a continuous\footnote{Stellar evolution codes use finite difference schemes to compute their solutions, so we do not mean ``continuous'' in the rigorous mathematical sense. The implication behind \textit{continuous} then is that there is a discrepancy in the values of the composition gradients on either side of the border 
that cannot be resolved with finer mesh resolution. Consequently, the discrepancy between the density and opacity values on either side of the border causes discontinuities (larger discrepancies) in both the radiative and adiabatic temperature gradients.} 
composition gradient across the convective border \citep{gabriel2014proper}. 
This occurs, for example, on the zero-age main sequence and at the red giant branch bump (due to thermohaline mixing). However, an issue arises once a composition discontinuity forms, given the difficulty of smoothing over such discrepancies numerically.
%
In this case, $\nabla_{\rm rad} = \nabla_{\rm ad}$ is only met on the convective side of the border. On the radiative side, we hold the inequality $\nabla_{\rm rad} < \nabla_{\rm ad}$. In the context of a convective core, using the sign-change algorithm with Schwarzschild and a composition discontinuity (or Ledoux with and without a discontinuity) results in the incorrect placement of the border. This typically inhibits the growth of the convective zone and so results in a less massive core. 

A more precise way to locate the convective border, whilst dealing with composition discontinuities, is to use a subgrid physics model. We define this approach as a ``direct search algorithm,'' depicted in the lower panel of Figure \ref{conv_regions}. Direct search algorithms begin with a candidate boundary, $j$. We then iteratively improve the resolution between $j$ and $j+1$ and consider how the discriminant would change if some intermediate, 
finer mesh shell, $k$, located between $j$ and $j+1$, were part of the convective region. If $k$ is unstable to convection under the relevant stability criterion, it is allocated to the well-mixed convective region\footnote{“well-mixed” here refers to a lack of stratification in the composition gradient. For our case, this just means the convective region where we assume instantaneous mixing. This is not the case where time dependent mixing is required-i.e. where the nuclear burning timescale is shorter than the mixing timescale.}. The process is repeated for mesh point $k+1$; if $k+1$ is still stable under these convective conditions, then $k$ is considered the formal boundary. Examples of direct search algorithms in use are the \textit{relaxation} method (or search for convective neutrality) defined in \citet{Lattanzio86} and implemented in the Monash code (and MONSTAR), as well as the ``predictive mixing'' and ``convective premixing'' schemes implemented in \texttt{MESA} (see \citealt{Paxton18instrument4} and \citealt{Paxton19instrument5}, respectively, for thorough discussions of these). Compared to the sign-change algorithm, direct search tends to result in larger convective zones. We see this in the more efficient TDU mixing episodes on the TP-AGB when using the \textit{relaxation} direct search algorithm, as the convective envelope is able to penetrate further into the internal radiative region (i.e. \citealt{Frost96}). 

Finally, a third type of boundary search algorithm exists, commonly known as convective overshoot \citep{shaviv1973convective} or convective boundary mixing \citep{Herwig00}. In the case of overshoot, we make manual adjustments to shift the border beyond mesh shell $j$. We define a particular distance, in units of pressure scale height ($\rm H_p$), over which material can permeate beyond the formal boundary, $j$. 
There are examples of overshoot algorithms in the literature using both diffusive and instantaneous treatments. The former include  
\citet{freytag1996hydrodynamical, schroder1997critical, bressan1981mass, Herwig00}, the latter \citet{Karakas10, Kamath12}. In comparison to sign-change and direct search, overshoot often leads to the largest 
growth of the convective core. 

 We know of several observed abundance anomalies that cannot be reproduced with standard stellar models (i.e. those that assume convection is the only form of mixing). For example, standard models incorrectly predict changes in the abundances of light elements--carbon, lithium and nitrogen--following the first dredge up event on the red giant branch \citep{carbon1982carbon, pilachowski1986abundance, gilroy1989carbon, kraft1994abundance, charbonnel1994clues, charbonnel1998many, gratton2000mixing, shetrone2019constraining, TayarJoyce22}. 
 As noted in \citet{schroder1997critical}, ``…the Achilles’ heel of modern evolutionary codes is (still) the very simple representation of convection.'' This is very much still relevant, some 25 years later. Not only is mixing in stellar interiors an extremely uncertain process, but we are also trying to approximate a 3D process in 1D. 
Some of the abundance discrepancies between theory and observation can be forced into consistency using a considerable amount of convective overshooting (e.g. \citealt{Joyce2015}). However, this amount of overshooting is applied \textit{ad hoc}, meaning it is not necessarily physical. More likely, it is a consequence of our current assumption that convection is the \textit{only} mechanism responsible for mixing in stellar interiors. There have been other mechanisms hypothesized in the literature including rotation--induced mixing \citep{palacios2003rotational}, atomic diffusion \citep{Bahcall95solar, Henney95effects, Gabriel97influence, Castellani97heavy, Chaboyer01heavy, bertelli18diffusion, liu19diffusion, semenova2020gaia}, internal gravity waves \citep{garcia1991li, denissenkov2003partial, rogers2013internal, varghese2022chemical} and thermohaline mixing \citep{Charbonnel07, Charbonnel10thermohaline1, Lagarde11thermohaline2, Lagarde12, Lattanzio15, Angelou15, Henkel2017, fraser2022characterizing}. The benefit, then, to overshoot is that it can be manipulated to absorb other 3D effects that are not included in our 1D models. For example, \citet{schroder1997critical} use overshoot to replicate giant star luminosities in $\zeta$ Aurigae systems. \citet{Kamath12} require a significant amount of overshoot (3$\rm H_p$, on top of the \textit{relaxation} algorithm) for their models to match observed C and O abundances in Magellanic Cloud clusters.

We note that a powerful constraint for the mixing profiles in stars arises from asteroseismology by using acoustic (p) and gravity (g) modes as probes. 
For example, \citet{pedersen2021internal} utilize the g mode to constrain the mixing profile of stars, given that g modes are quite sensitive to the core-boundary layer\footnote{g mode period spacing patterns should be constant for observations of non-rotating stars with homogeneous chemical composition and change only as mixing ensues in the stellar interior.}. They calculate mixing profiles for 26 stars with initial masses between 3 and 10\(M_\odot\), based on observed g mode period spacing patterns, and compare these profiles against a grid of theoretically derived mixing profiles. These theoretical profiles contain a variety of the common mixing assumptions used in the literature. They find that $\approx 65$\% of the sample profiles are better modelled using an instantaneous approximation for overshoot, as opposed to diffusive, at the core boundary layer. This method of course contains its own uncertainties and approximations, and we are still comparing 1D phenomenological prescriptions against asteroseismic oscillations from 3D stars. However, this method is one of few with which we can infer what goes in underneath the stellar surface and the initial results are promising. 

With our current understanding, it perhaps makes the most sense to consider models utilizing sign-change algorithms as conservative estimates of $M_{\rm mas}$, those using convective overshooting as upper limits of $M_{\rm mas}$, and those using direct search algorithms somewhere in the middle. With this in mind, the appearance of overshooting in models of Figure \ref{fig:comp} does not predict whether $M_{\rm mas}$ falls within a particular mass range, just that it results in a larger core mass and thus lower $M_{\rm mas}$. In both the \citet{Siess07evolutionII} and \citet{Eldridge04prog} models, the inclusion of overshoot lowers the $M_{\rm mas}$ quantity by $\approx 1 - 1.5$\(M_\odot\). However, for \citet{Siess07evolutionII} (using STAREVOL), this transition is from $M_{\rm mas}=10.93$\(M_\odot\) (no core overshoot) to 8.83\(M_\odot\) (with core overshoot) at $Z=0.02$. For \citet{Eldridge04prog} (using STARS), the transition is from $M_{\rm mas}=9.5$\(M_\odot\) (no core overshoot) to 7\(M_\odot\) (with core overshoot) at $Z=0.02$. Both codes decrease $M_{\rm mas}$ by $>$2\(M_\odot\), but \citet{Siess07evolutionII} is shifted ``into'' the common range with overshoot, whereas \citet{Eldridge04prog} is shifted ``out.'' Although the algorithms that we use to define convective borders contribute a large uncertainty, there is clearly more at play here with $M_{\rm mas}$. Both \citet{Siess07evolutionII} and \citet{Eldridge04prog} use different overshoot prescriptions (\citealt{freytag1996hydrodynamical, Herwig00} for the former, \citealt{schroder1997critical} for the latter), but it is unlikely that this makes a significant difference given that the inclusion of either prescription lowers the core masses by a similar magnitude (30\% and 21\%). This suggests that there is either a combination of other physical parameters that accumulate to make a significant difference and/or that the numerics of the software instruments themselves have a more substantial dependence on the core mass beyond the treatment of convective boundaries. 
In the next few subsections, we discuss other approximations that will likely make lesser, but non-negligible, contributions to variations in $M_{\rm mas}$ between evolution codes. We do not discuss here the impact software architecture can enact on evolution results (even when implementing the same input physics) given that this was the focus of Paper 1 \citep{BridgingIcinquegrana}. On this topic, we refer the interested reader to the following works focused on the inter-comparison of particular stellar evolution tools:
\begin{itemize}
\item[-] \citet{Paxton10instrument1}, compares \texttt{MESA} against BaSTI (a Bag of Stellar Tracks and Isochrones; \citealt{Pietrinferni04large, Hidalgo18updated, Pietrinferni21updated, Salaris22updated}), FRANEC (the Frascati Raphson Newton Evolutionary Code; \citealt{Chieffi98evolution, Limongi06nucleosynthesis}), DSEP (the Dartmouth Stellar Evolution Program; \citealt{Dotter07acs}), GARSTEC (the Garching Stellar Evolution Code; \citealt{Weiss08garstec}) and EVOL. 

\item[-] \citet{Martins13comparison}, compares \texttt{MESA} and STAREVOL against the Geneva stellar evolution code (\citealt{Eggenberger08geneva}), STERN, PARSEC and FRANEC. 

\item[-] \citet{Sukhbold14compactness}, compares \texttt{MESA} and KEPLER. 
\item[-] \citet{Jones15codecomp}, compares GENEC, KEPLER and \texttt{MESA}. 
\item[-] \citet{Joyce2015} compares DSEP against BaSTI, YREC, PARSEC, and others. 
\item[-] \citet{Aguirre20aarhus}, compares nine stellar evolution codes: BaSTI, GARSTEC, \texttt{MESA}, MONSTAR, YREC (the Yale Rotating Stellar Evolution Code; \citealt{Demarque08yrec}), ASTEC (the Aarhus STellar Evolution Code; \citealt{Dalsgaard08astec}), CESTAM (Code d’Evolution Stellaire Adaptatif et Modulaire; \citealt{Morel08cesam, marques2013seismic, deal2018impacts}), LPCODE (La Plata Observatory, \citealt{althaus03lpcode}) and YaPSI (\citealt{skumanich72yapsi}).  

\item[-] \citet{Agrawal22explaining} compares Geneva, \texttt{MESA}, PARSEC, BPASS (Binary Population and Spectral Synthesis; \citealt{eldridge2017binary}), BoOST (Bonn Optimized Stellar Tracks; \citealt{szecsi2022bonn}). 
\item[-] \citet{Campilho22atomic} compares \texttt{MESA} against the Montreal/Montpellier stellar evolution code (\citealt{turcotte1998consistent, richer2000evolution}) and with CESTAM.
\item[-] Paper I \citet{BridgingIcinquegrana}, compares \texttt{MESA} against the Monash stellar evolution code. 
\end{itemize}

\subsection{Treatment of convection}

Convection is an extremely uncertain process to model in 1D. Whilst we were previously concerned with the borders that define convective regions, here we focus on the actual convective treatment. To model a star, we need to solve the equations of stellar structure and calculate the general temperature gradient, $\nabla$ (defined previoulsy in Equation. \ref{generaltemp}). In regions of dynamic stability, energy is transported via radiation. Thus, $\nabla$ comprises radiative flux contributions: 
\begin{equation}
    F_{\rm rad} = \frac{4 a c G}{3} \frac{T^4 m}{\kappa P r^2} \nabla,
\end{equation}
where $a$ is the radiation density constant and $c, G, T, m, P, r$ have their usual meanings. In regions of dynamic instability, some to all of the energy flux is transported via convection. 
%
In the superadiabatic case, the general temperature gradient, $\nabla$, comprises both radiative and convective flux contributions. The latter requires some approximation to compute in 1D; all works compared here utilize the Mixing Length Theory (MLT) of convection to do so (though some other stellar evolution codes do include alternatives, e.g. the Full Spectrum Turbulence model of \citealt{Canuto91, Canuto96}, we do not examine results from these here).
The MLT assumes that a mass "blob" will travel some distance, $\lambda$, before dissolving into its surroundings. Under this basic approximation, we can derive the following expression for $F_{\rm conv}$, 
\begin{equation}\label{Fconv}
F_{\rm conv} = \frac{1}{2} \rho v c_p T \frac{\lambda}{\rm H_p} (\nabla - \nabla_{\rm ad}),
\end{equation}
where $\rho$, $v$ and $T$ have their usual meanings, $c_p$ is the specific heat, $\nabla$ and $\nabla_{\rm ad}$ are the general (or true) and adiabatic temperature gradients. $\lambda$ is the mean free path of the gas element, $\rm H_p$ is the pressure scale height, and so $\frac{\lambda}{\rm H_p}$ is the mixing length parameter, $\alpha_{\rm MLT}$. Using this definition, we can approximate the true temperature gradient $\nabla$ in a given dynamically unstable region. For clarity, we also define the convective efficiency, $\Gamma$, which from \citet{Kippenhahn12} is given by:
\begin{equation}\label{eq:gamma}
\Gamma = \frac{4}{3} \frac{F_{\rm conv}A}{\lambda}.
\end{equation}
Here, $A$ is the cross section of the mass "blob" and $\lambda$ is the total energy loss. Given that the MLT is a phenomenological--rather than physical—model, $\alpha_{\rm MLT}$ must be calibrated for each stellar evolution code and ideally, each set of input physics. This calibration is typically performed against the Sun (e.g. \citealt{Cinquegrana22solarcal}), although arguments have been made that the Sun is not always the best choice \citep{Joyce18not, Joyce18class}. The MLT was originally established by \citet{Prandtl25} but first developed for stellar interiors by \citet{Bohm1958}. Since then, it has been modified for specialization in optically thin \citep{Henyey65mlt} and optically thick \citep{Cox68} regimes. It is important to note, however, that the MLT only needs to be solved in regions of significant (but not too significant) super adiabaticity. Here, $\Gamma$ (from Equation. \ref{eq:gamma}) is low and the true temperature gradient lies between $\nabla_{\rm rad}$ and $\nabla_{\rm ad}$. This occurs in the outer layers of giant star envelopes, where the temperature gradient must increase significantly to transport heat through the extremely opaque material. Convection in the central regions of stars is technically superadiabatic, but it is so efficient ($\Gamma \rightarrow \infty$) that we approximate $\nabla = \nabla_{\rm ad}$. Thus, there is no need to solve the MLT equations. On the other hand, if the region is \textit{too} superadiabatic (e.g. the stellar photosphere, where $\Gamma \rightarrow 0$), then convective transport is so inefficient that radiation is carrying almost all the flux. In that case, we can approximate $\nabla = \nabla_{\rm rad}$ (see \citealt{JoyceTayar2023} for thorough coverage of this topic).

Although the MLT is not used to model core convection, the use of various MLT formulations (e.g. \citealt{Bohm1958, Henyey65mlt, Cox68, bohm1971convective, mihalas1978two, mihalas1978stellar, kurucz1979model}) and calibration methods for $\alpha_{\rm MLT}$ will have an indirect impact on the size of the core. Suppression of the surface convective flux---whether due to a small $\alpha_{\rm MLT}$, magnetism, certain atmospheric boundary conditions, or otherwise---leads to a corresponding structural re-alignment of the star that propagates to the inner-most point of the model. In essence, we are adjusting the boundary conditions of the coupled equations of stellar structure describing our model. The knock-on structural effects of small $\alpha_\text{MLT}$ is demonstrated in \citet{JoyceTayar2023} for the case of a 1\(M_\odot\), \(Z_\odot\) model. On the main sequence, stars with initial masses between 0.5 and 1.2\(M_\odot\) should have radiative cores with convective envelopes. However, in the case where $\alpha_{\rm MLT}$ is reduced below 0.5$\rm H_p$, the impediment to the expulsion of convective flux is so great that the 1\(M_\odot\) model compensates by developing a convective core.
 
\subsection{Semiconvection}

At this point, we have discussed the fact that our treatment of convection and our placement of convective boundaries will have an impact on our H exhausted core (and thus, $M_{\rm mas}$). The aim of this section is to consider the effects of a more nuanced mixing process, \textit{semiconvection}. Semiconvection, and its place within the context of stellar interiors, has been recognized by the astronomical community since the 1950s \citep{Tayler54, schwarzschild1958evolution}. Since then, it is not the actual presence of these zones, but the extent of mixing that occurs inside of them, that has been hotly debated in the literature. In this section, we discuss our understanding of this process, what light it shines on semiconvective mixing efficiency, and lastly the potential impact this enacts on $M_{\rm mas}$. 

Semiconvection is a type of double-diffusive instability\footnote{We refer the reader to \citet{zaussinger2017semi, garaud2018double, garaud2021double} for reviews on double-diffusive instabilities relevant to stellar interiors}, known as the oscillatory double-diffusive instability (ODDC), that occurs in regions that are thermally unstable (according to Eq. \ref{schw}) but have a stable ($\mu < 0$) composition gradient (according to Eq. \ref{ledoux})\footnote{Thermohaline convection is another form of a double-diffusive instability (known in geophysics as the salt fingering instability). In contrast to semiconvection, thermohaline mixing occurs in thermally stable regions (i.e. where radiative transport dominates) with an unstable ($\mu < 0$) composition gradient \citep{garaud2021double}}:
\begin{equation}
\nabla_{\rm ad} < \nabla_{\rm rad} < \nabla_{\rm ad} + \frac{\phi}{\delta} \nabla_{\mu}. 
\end{equation} 
Although stellar semiconvection has been discussed since the 1950s, it was not classified as ODDC until close to a decade later \citep{kato1966overstable, spiegel1969semiconvection}. 

Our definition of semiconvection invites two main questions. Firstly, how do compositional gradients form in stars? The two most common examples of stable $\mu$ gradients are during core H burning in massive stars (example of a receding convective core) and during core He burning in low and intermediate mass stars (example of a growing convective core). The former situation, for example, arises from the fact that the opacity in main sequence massive stars is dominated by electron scattering, which is directly proportional to H content: $\kappa \sim (1+\rm X_H)$. Thus, as the H content inside the convective core begins to decline, a higher percentage of the energy is now transported via radiation. Consequently, the boundary of the convective core begins to retreat. This leaves shells of H layers of increasing H content outwards, as each shell is successively less nuclearly processed than the previous.
%
Secondly, we might then also ask what occurs once the conditions for semiconvection are met? The physics of double diffusive convection is fairly well understood, however earth-based experiments are often performed with high Prandtl number fluids (e.g. salt water, with $Pr \sim 7$) and so do not directly translate to stellar interiors, where Prandtl numbers and diffusivity ratios are much lower ($Pr$ and $\tau << 1$). The best chance we have at studying ODDC in this context is through 3D numerical simulations. For example, one of the most well-known studies is by \citet{rosenblum2011turbulent}, whose calculations extend down to $Pr = 0.3$ and $\tau = 0.3$\footnote{They note that achieving lower values is possible, but the computational expense of the simulation increases significantly.}. In their work, \citet{rosenblum2011turbulent} identified the behaviour of two different types of ODDC that can arise in astrophysical conditions, homogeneous and layered, the latter of which occurs as the inverse density ratio, 
\begin{equation}
    \rm R^{-1}_0 = \frac{\nabla_{\mu}}{\nabla - \nabla_{\rm ad}}
\end{equation}
decreases below $\rm R^{-1}_0 < 1.35$. Layered convection, as the name suggests, forms in a series of many thin layers. 
These layers consist of both efficient, overturning convection and stagnant zones, stabilised by the density gradient, where transport occurs via molecular diffusion (which is much slower than convection). So, it is the stagnant zones that slow down the transport in these regions. For further discussion of the layering phenomenon, see \citet{spruit1992rate, radko2003mechanism, spruit2013semiconvection, zaussinger2013semiconvection}.

\citet{moll2016new} further showed that, given this layering results from the gamma instability \citep{radko2003mechanism}, layered convection can further be split into two types, based on how the layers form. A crucial aspect to this discussion, however, is to consider how energy is transported through each type of semiconvective zone. \citet{moll2016new} confirm that the two different types of ODDC lead to very different energy transport efficiencies, with the fluxes in non-layered semiconvective zones no larger than those of conduction or molecular diffusion. Therefore, it is important to know what type of semiconvective zone we are modelling. \citet{moore2016main} find that semiconvective regions that are close to fully convective zones, such as in a convective core, are always layered. 

We are most interested in the layered semiconvection which occurs adjacent to the fully convective core, that can potentially impact the convective core size. Layered convection is suspected to be significantly more efficient at transporting energy flux than the diffusive non-layered zones, however the extent of that efficiency is contested. Following the findings of \citet{rosenblum2011turbulent}, \citet{moore2016main} implemented the recent layered semiconvective approximation of \citet{wood2013new} into \texttt{MESA}. They find that the resultant mixing---which occurs in layered semiconvective zones—--is heavily dependent on the height of the layer. In fact, there is a critical layer height above which mixing is so efficient that it is comparable to using the Schwarzschild criterion and below which mixing is so inefficient that the model evolves the same as if they had used the Ledoux criterion. Importantly though, this critical layer height is appreciably \textit{smaller} than the \textit{predicted minimum height} for these layers. This implies that any layered convection occurring in stars is likely to have layer heights above this threshold and so the mixing is likely to be quite efficient. The results of \citet{moore2016main} therefore suggest that the magnitude of the convective core produced is comparable to that using the Schwarzschild criterion. 

The current semiconvective prescriptions in use (of which there are many) reflect a variety of physical processes and approximations. The ideal implementation, that best reflects our current understanding, is one that can switch between using both non-layered and layered prescriptions for semiconvection based on the model. An example of this is provided by \citet{mirouh2012new, wood2013new, moll2016new}. 
We lack information on the specific semiconvective prescriptions used in the papers to which we compare here, but we can get an approximate idea of the maximum potential difference between the models with the following: if the mixing in semiconvective regions is so efficient that the composition gradients are erased quickly and the region itself is absorbed into the convective core, then the maximum contribution semiconvection can add to the core mass is the difference between the core masses predicted by Ledoux and Schwarzschild criteria. 
If, by the other extreme, the mixing is completely inefficient, then the convective core will remain the magnitude predicted by the Ledoux criterion. So, the difference in the convective core magnitude between a model that uses the Ledoux criterion with no semiconvection prescription (no mixing), one that uses the Schwarzschild criterion (full convective mixing) and one that uses the Ledoux criterion \textit{with} a semiconvective prescription (partial mixing) will be at most the difference in mass between the predictions of Schwarzschild vs Ledoux. 
The results of \citet{moore2016main} suggest that those using the Schwarzschild criterion will produce the largest (and hence most realistic) core masses.

With a proper implementation of ODDC semiconvection in their 3D simulations, \citet{Anders22schwarzschild} find that the Ledoux and Schwarzschild criteria are equivalent on evolutionary timescales. The convection zone produced using the Ledoux criterion is initially smaller than the zone predicted by Schwarzschild, but the convection zone produced using the Ledoux criterion grows via entrainment\footnote{The incorporation of entrainment into 1D stellar models has been investigated by \citet{staritsin2013turbulent} and \citet{ scott2021convective}. In the 8\(M_\odot\) models of \citet{scott2021convective} in particular, significantly more massive He cores are produced using their entrainment algorithm over the traditional overshooting prescriptions, such as those discussed in \S~\ref{CB}. The advantage to their algorithm is that the penetrability of the boundary between convective and radiative regions is not treated as static; both in terms of  evolutionary phase and initial mass.}, eventually to the same size as that predicted by the Schwarschild criterion. We note however, this was achieved by properly modelling ODDC semiconvection. For 1D simulations, they recommend using the Schwarzschild criterion.

\begin{figure} 
	\includegraphics[width=8cm]{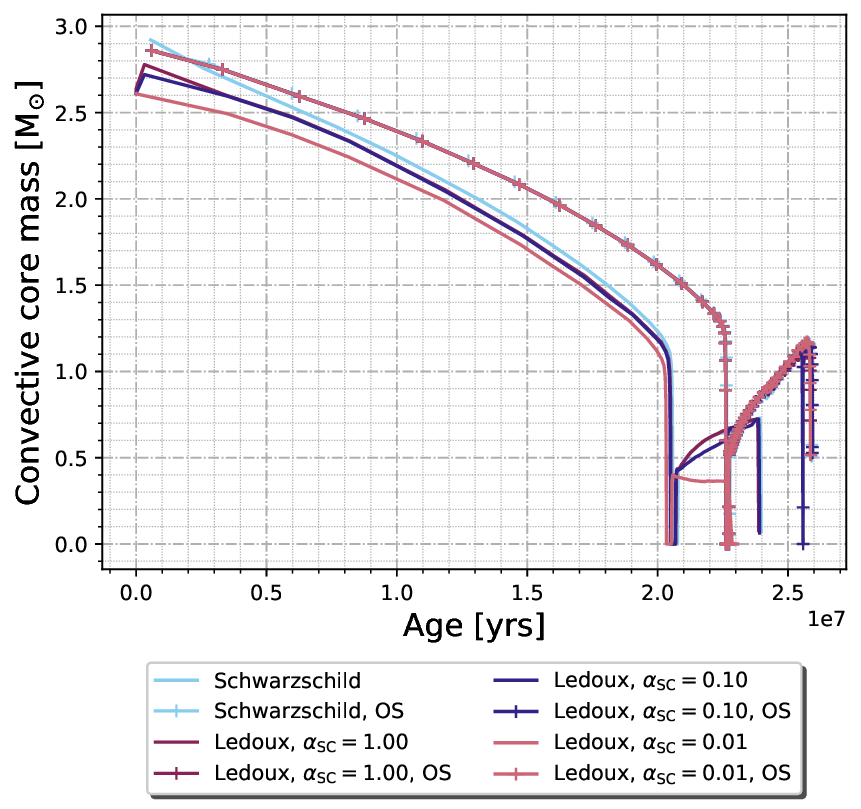}
    \caption{The progression of the convective core growth in 9.45\(M_\odot\) models with $Z=0.02$ from the zero-age main sequence to the end of core He burning (TACHeB). The models use either the Schwarzschild or Ledoux stability criterion, with varying levels of semi convective efficiency and convective overshoot.}
    \label{fig:led_schw}
\end{figure}

To provide a quantitative description of the impact of these choices on convection zone size, in Figure \ref{fig:led_schw} we show the progression of the convective core growth of 9.45\(M_\odot\) models with $Z=0.02$ from the ZAMS to TACHeB when using the Ledoux or Schwarzschild stability criterion, and both with and without convective overshoot. For the models without convective overshooting (solid lines), there is an observable discrepancy between all the convective core masses on the main sequence, but this difference becomes most pronounced when the model undergoes core He burning. All models converge to approximately the same core mass at TACHeB ($\approx 0.7$\(M_\odot\)), except for the Ledoux model with the lowest semiconvective efficiency (as $\alpha_{\rm SC} \rightarrow 0$, the convective border reduces to that predicted by the Ledoux criteria alone). The convective core of this model is $\approx 0.4$\(M_\odot\) at TACHeB, instead. The difference in the convective core masses with overshoot are much less significant (also found in \citealt{kaiser2020relative}, to which we refer the reader for a detailed discussion into the variation of convective zone growth with different stability criteria and overshooting efficiencies). 

\subsection{Carbon fusion reaction rates}

Nuclear reaction rates, in general, comprise a major role in stellar modelling uncertainties \citep[e.g.][]{Lugaro04reaction, Karakas06, Herwig06nuclear, Izzard07, Van08}, particularly for later burning stages as found by \citet{fields2018impact}. However, the uncertainties of the C burning rate, $^{12}$C$+$$^{12}$C, are particularly important for stellar structure. In fact, \citet{fields2018impact} test the impact of 665 temperature-dependent reaction rate uncertainties (from the STARLIB library, \citealt{Sallaska13starlib}) on various stellar modeling quantities. At core C and Ne depletion, they deem the $^{12}$C($^{12}$C, p)$^{23}$Na rate the key reaction that governs the size of the ONe core mass, which makes this reaction pertinent to our discussion. 

C burning occurs for central temperatures greater than $T_c > 0.5 \times 10^8$K, where $^{12}$C particles can fuse together to produce the compound nucleus, $^{24}$Mg, which can then decay via three channels: $^{12}$C($^{12}$C, $\alpha$)$^{20}$Ne, $^{12}$C($^{12}$C, p)$^{23}$Na or $^{12}$C($^{12}$C, n)$^{23}$Mg. The astrophysical temperature range for quiescent C burning ($\approx$ 0.8 to 1.2$\times 10^8$K; \citealt{tumino2021trojan}) corresponds to energies of $\approx$ 1-3MeV. As a result, direct measurements of the $^{12}$C$+$$^{12}$C reaction rates are extremely difficult to perform due to the increasingly small particle cross sections that accompany low energies (see \citealt{monpribat2022new} for a thorough historical overview of the $^{12}$C$+$$^{12}$C problem and experiments). In practice, then, C-fusion rates are generally measured at higher energies of $>2$MeV and extrapolated down to sub-coulomb energies. The major issue that arises with this method is the possible presence of unknown resonance \citep{spillane2007c} or hindrance \citep{jiang2007expectations} effects that are suggested to exist around the Gamow window, $\approx 1.5$MeV. The two phenomena have opposing consequences, with the former predicting our extrapolated rates are too slow and the latter, too fast. 

Faster (slower) carbon burning rates result in lower (higher) core carbon ignition temperatures \citep{pignatari201212C, bennett2012effect, chen2014dependence, fields2018impact}. For example, \citet{monpribat2022new} present two new sets of $^{12}$C$+$$^{12}$C rates---one based on the hindrance model, the other including a resonance contribution---and compare how these new rates impact the evolution of 12\(M_\odot\) and 25\(M_\odot\) models as compared to the standard \citet{Caughlan88} $^{12}$C$+$$^{12}$C rate. They find that the models using the hindrance model (and thus a lower rate) ignite carbon at a core temperature 10\% greater than in the models using the faster rates. Higher carbon burning temperatures suggest that $M_{\rm mas}$ should increase with slower rates. Similarly, \citet{straniero2016we} study the minimum mass for a WD, $M_{\rm up}$. When they incorporate resonance contributions at 1.4MeV to CF88, $M_{\rm up}$ is driven down by 2\(M_\odot\), which suggests that $M_{\rm mas}$ would be reduced by a similar amount. Faster carbon fusion rates also appear to have a non-negligible impact on neutron capture nucleosynthesis \citep{bennett2012effect, pignatari201212C}. \citet{bennett2012effect} find that when including a resonance contribution at 1.5MeV, the dominant neutron source changes from $^{22}$Ne($\alpha$, n)$^{25}$Mg to $^{13}$C($\alpha$, n)$^{16}$O. This results in an order of 2 magnitude increase in their \textit{s}-process yields. Further testing of resonance and hindrance effects in stellar models are performed by \citet{bravo2011type, gasques2007implications}. 

The bulk reaction rates in the STARS, STAREVOL and EVOL evolution codes come from NACRE \citep{angulo1999compilation, xu2013nacre}, \texttt{MESA} utilizes JINA REACLIB \citep{Cyburt10} and MONSTAR uses CF88 \citep{Caughlan88} rates for carbon burning. However, NACRE and JINA REACLIB---together with the other most common reaction rate library utilized in stellar modelling, STARLIB \citep{Sallaska13starlib}---defer to CF88 for the $^{12}$C$+$$^{12}$C rates. Thus from the information we have available, we expect all works compared here to use the same $^{12}$C$+$$^{12}$C rates. Consequently, variation in $M_{\rm mas}$ here is not due to differing carbon fusion rates, but the $^{12}$C$+$$^{12}$C rates remain a significant source of uncertainty in general. 

\subsection{Defining the magnitude of the core mass}

When comparing our results to the literature, it is also important to consider the definition each group uses to classify a Fe CCSNe progenitor. In \citet{Siess07evolutionII}, \citet{Doherty15super}, and this work, a massive star is defined as one with an Oxygen-Neon, or ONe, core mass $\geq$ 1.37\(M_\odot\) at core C depletion. \citet{Eldridge04prog, Poelarends08supernova} classify massive stars as those with CO core masses greater than the Chandrasekhar mass (1.38\(M_\odot\); \citealt{chandrasekhar1935highly}) following the second dredge up. \citet{Ibeling13} require both Si ignition and that $M_{\rm ONe} > 1.38$\(M_\odot\). 

Following core He burning, the H--exhausted core mass can be reduced by two mixing processes: the second dredge up (SDU) and the ``dredge out'' (see \citealt{Doherty17}). The evolutionary timing of C ignition and the second dredge up is dependent on the initial stellar mass (greater $T_c$ for greater $M_i$), where core C ignites earlier (prior or during SDU) in more massive stars. Some great examples of the progression of C burning against the backdrop of the SDU (or dredge out, in the most massive models) are shown in Figure 10 of \citet{Siess06evolutionI}, for $9 \leq M_i < 11.3$, and Figure 4 from \citet{jones2013advanced} which covers the mass range $8.2 \leq M_i < 12$. 

Given these additional mixing processes, there could be small reductions in the core mass if our models were permitted to evolve beyond their prescribed termination condition. The other consideration here is that the central C mass fraction is not completely indicative of the termination of core C burning. C flames can propagate off centre and still burn even when the central reserve of carbon has been exhausted. However, the stopping condition utilised in \citet{Ibeling13} takes all these factors into account, yet on average our values differ from theirs by $\sim 0.5$\(M_\odot\).
%
We re-iterate that there are likely many other factors that will hold some influence over $M_{\rm mas}$: mass loss, rotation, 
magnetism, other (non-convective) forms of mixing. Further, the actual definitions of core boundaries (and thus magnitudes) differ between evolution codes. Regardless of these non-trivial uncertainties, however, we still observe the same general qualitative trend for $M_{\rm mas}$ with metallicity across the literature. 

\section{Observational constraints}\label{sec:observation}

Having examined the extent of uncertainty in our modelling choices, it is worthwhile to consider observational constrains on $M_{\rm mas}$. In this section, we review some common methods for measuring the progenitor mass of CCSNe and then probe an example from the literature of how these observations are used to constrain $M_{\rm mas}$.

\subsection{Measuring the progenitor mass of an observed CCSNe}

The best known method for determining the initial mass of an observed SN is to identify the progenitor in images taken prior to the event, as performed in \citet{Smartt09death, smartt2015observational}. These studies collate a sample of pre-explosion images and obtain luminosities (or upper luminosity limits where the object is not explicitly detectable). These are compared to theoretical models, generated with the STARS evolution code in \citet{Smartt09death} and both STARS and KEPLER in \citet{smartt2015observational}, which provides a relation between progenitor initial mass and final luminosity. Other studies that utilize this method include \citet{elias2009progenitor, fraser2011sn, van2011progenitor} and \citet{maund2011yellow}. 

However, the utility of this method is limited given that the SNe events need to be relatively nearby and have an image taken at the right time. \citet{spiro2014low} determine progenitor masses of low luminous SNe IIP by instead comparing theoretical light curves calculated with hydrodynamical models against their observational counterparts. They determine that low--luminous SNe descend from $10-15$\(M_\odot\) type stars, which is in agreement with direct image predictions (see also \citealt{utrobin2007optimal, dessart2010determining, roy2011sn, bersten2011hydrodynamical, tomasella2018sne, martinez2019mass} and \citealt{limongi2020hydrodynamical}). 

\citet{jerkstrand2012progenitor} use spectral modelling as SNe cool into their nebular phase to derive progenitor mass estimates\footnote{Thesis available at: https://arxiv.org/pdf/1112.4659.pdf}. They compare emission lines (particularly C, O, Ne, Na, Mg, Si and S), monitored over a period of 140-700 days, to nucleosynthesis models to predict the progenitor mass given that the composition of nucleosynthesis products is dependent on initial mass\footnote{Amongst other variables not discussed here, such as metallicity}. In the case of SN 2004et, the \citet{jerkstrand2012progenitor} value of 15\(M_\odot\) is a good match to what is predicted by pre-explosion imaging: 14\(M_\odot\). However, this is somewhat in tension with progenitor mass estimates calculated with hydrodynamic models, which typically fall at or above $25$\(M_\odot\). 

\citet{gogarten2009ngc} utilise the fact that the surrounding population of stars immediately surrounding a SN event should share the same age and chemical composition as the progenitor. They fit photometric data to theoretical stellar evolution models from \citet{girardi2002theoretical} and \citet{marigo2008evolution} to obtain the most suitable ages and metallicities that correspond to the colors and magnitudes observed 
 (the same method as in \citealt{williams2008acs} and \citealt{gogarten2009acs}). Once they have determined the age of the population, \citet{gogarten2009ngc} identify the masses of surrounding stars at the main sequence turn off and early subgiant branch. This provides upper and lower mass limits for the main sequence mass of the progenitor; however, it does not determine the phase of evolution the progenitor was in when it exploded (see also \citealt{jennings2012supernova} and \citealt{jennings2014supernova}). 

\subsection{Deriving an observed \texorpdfstring{$M_{\rm mas}$}{Mmas}}

After the progenitor masses of observed supernovae have been measured, the next step is to quantify an observed lower mass limit. An example of one such calculation is provided by \citet{botticella2012comparison}, who use the relationship between the birth (star formation rates; hereafter SFR) and death rates (CCSNe rates) of massive stars. 
Observed CCSNe rates (reviewed in \citealt{van1991galactic}) were first measured by \citet{zwicky1938frequency} and subsequently in e.g. \citet{van1991galactic, van1987supernova, cappellaro1996rate, botticella2008supernova, li2011nearby}. CCSNe events are often used as indicators of instantaneous star formation rates in galaxies due to the brief lifetimes of massive stars. Using an IMF, this can then be extrapolated to estimate a star formation rate (SFR) for the entire mass regime. SFR and CCSNe rates are linked by the following relationship \citep{botticella2008supernova}, 
\begin{equation}\label{CCrate}
    r^{\rm CC}(z) = \frac{\int_{m_l^{\rm CC}}^{m_u^{\rm CC}}\phi(m){\rm d}m}{\int_{m_L}^{m_U}m\phi(m){\rm d}m} \times \psi(z)
\end{equation}
where $r^{\rm CC}(z)$ is the CCSNe rate, $\phi(m)$ is the IMF, $m_L$ and $m_U$ are the lower and upper mass bounds for a population, $\psi(z)$ is the SFR and $m_l^{\rm CC}$ and $m_u^{\rm CC}$ are the lower and upper mass bounds for CCSNe progenitors. In this definition, $M_{\rm mas} = m_l^{\rm CC}$.

\citet{botticella2012comparison} measure the SFR through UV and H$\alpha$ emission tracers and the observed CCSNe rate for the same galaxy within the local volume. They measure an $M_{\rm mas}$ value of approximately $8 \pm 1$\(M_\odot\) ($6 \pm 1$\(M_\odot\)) where the SFR is measured using the far ultraviolet (H$\alpha$) luminosities. 

Other methods of constraining $M_{\rm mas}$ can be found in \citet{Smartt09death, smartt2015observational}, for example. With 20 derived SNe progenitor masses in their sample, they use a maximum likelihood analysis to derive an average lower value of $M_{\rm mas}=8.5^{+1.0}_{-1.5}$, $M_{\rm mas}=9.5^{+0.5}_{-2}$ and $M_{\rm mas}=10^{+0.5}_{-1.5}$\(M_\odot\), depending on the theoretical models used to derive the ZAMS masses. We can also switch our perspective to the other end of the mass spectrum, by identifying the maximum mass of a WD. \citet{dobbie2006new} calculate a WD upper progenitor mass limit of $6.8 - 8.6 M_\odot$. 

Unfortunately, all observed estimates for $M_{\rm mas}$ are quoted without dependence on [Fe/H]. 
Nonetheless, these observations do still contribute to constraints on the behavior of $M_{\rm mas}$ as a function of mass, as shown in Figure \ref{fig:comp}. However, observational techniques are subject to their own uncertainties and biases. One source of bias is due to the fact that SNe are  observed by chance, rather than in systematic, intentionally designed observational campaigns. It is also quite easy to miss less luminous events. Consequently, there is a large discrepancy between observational and theoretical SNe rates (with theory overpredicting observed rates, see discussions in \citealt{van1991supernova, van1993rare, botticella2008supernova, horiuchi2011cosmic}). Observed SFRs share this uncertainty; in the example above of \citet{botticella2012comparison}, there was a difference of 2\(M_\odot\) in calculated $M_{\rm mas}$ depending on whether the SFR was measured using far ultra violet or H$\alpha$ luminosities. There are likewise systematic errors when measuring progenitor masses. \citet{davies2018initial} attempt to address one of these: namely, that when using preSN photometry to measure initial masses, the process often does not account for changes in the bolometric correction between stellar phases. Many of the observational studies mentioned above need to utilize some form of theoretical modelling. \citet{smartt2015observational} quote a potential 1\(M_\odot\) difference in the $M_{\rm mas}$ lower limit based on whether the STARS or KEPLER theoretical models were used to calculate the ZAMS masses. Thus, our theoretical modelling uncertainties propagate into the observed $M_{\rm mas}$ uncertainties themselves. 

\section{Implications for the galactic community}\label{sec:implications}

\begin{figure} 
	\includegraphics[width=8cm]{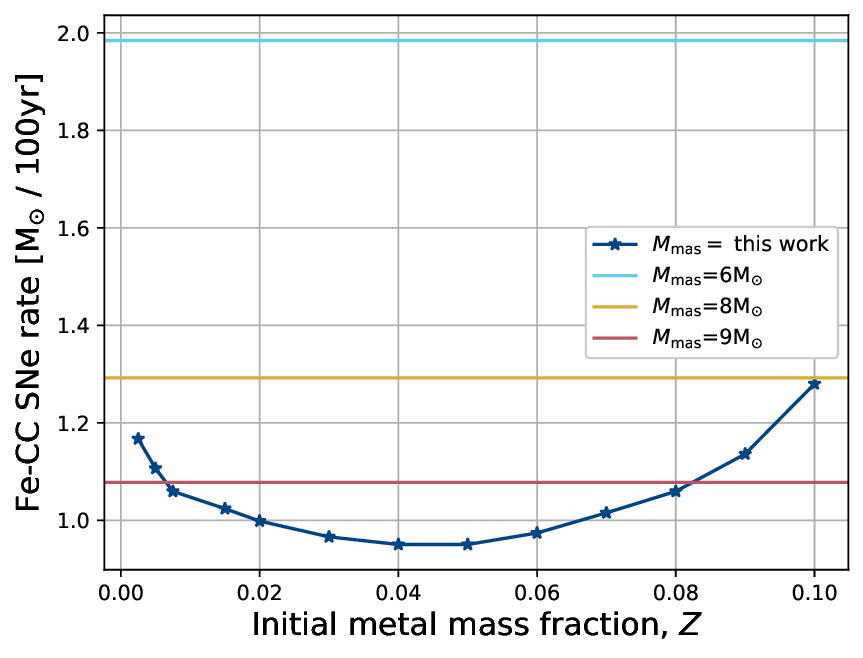}
    \caption{Here, we show estimates for Fe CCSNe rates (calculated using Equation. \ref{CCrate}) for metallicity independent $M_{\rm mas}$ values (6, 8 and 9\(M_\odot\)) and the metallicity dependent $M_{\rm mas}$ values derived in this work. This figure is produced using the following values and prescriptions: we use a standard Salpeter IMF \citep{Salpeter55} of $\phi(m)=m^{-2.35}$ for a galactic population with masses between 0.1\(M_\odot\) ($m_L$) and 100\(M_\odot\) ($m_U$). We set $m_u^{\rm CC}=40$\(M_\odot\) and $m_l^{\rm CC}=M_{\rm mas}$ as a function of $Z$ and use a value of $\psi(z)=1.9$\(M_\odot\)yr$^{-1}$ (based on estimates of the Milky Way SFR; \citealt{chomiuk2011toward, kennicutt2012star}). As stated in the text, we caution the reader that this is purely an example of how using metallicity dependent $M_{\rm mas}$ can influence the Fe CCSNe rate, using rough estimates for the other values in Equation. \ref{CCrate}. These rates need to be considered more thoroughly before being used for quantitative purposes.}
    \label{fig:sne_rate}
\end{figure}

In light of the increasingly rich data climate of the modern observational era, we also consider the potential impact of a theoretically derived $M_{\rm mas}$ for the galactic community. From Equation \ref{CCrate}, we can use an observed SFR together with our $M_{\rm mas}$ in Equation \ref{CCrate} to calculate a CCSNe rate and vice versa. We show, for example, how the CCSNe rate varies with $M_{\rm mas}$ in Figure \ref{fig:sne_rate}. 
To produce Figure \ref{fig:sne_rate}, we use a standard Salpeter IMF \citep{Salpeter55} of $\phi(m)=m^{-2.35}$ for a galactic population with masses between 0.1\(M_\odot\) ($m_L$) and 100\(M_\odot\) ($m_U$). We set $m_u^{\rm CC}=40$\(M_\odot\) and $m_l^{\rm CC}=M_{\rm mas}$ as a function of $Z$ and use a value of $\psi(z)=1.9$\(M_\odot\)yr$^{-1}$ (roughly appropriate for Milky Way star formation; \citealt{chomiuk2011toward, kennicutt2012star}). We emphasize here that the example presented in this Figure was calculated using rough estimates for the other values in Equation \ref{CCrate} to serve purely as an example of how using metallicity--dependent $M_{\rm mas}$ can influence the Fe CCSNe rate. 
In reality, there are large uncertainties, and the inadequacy of using one constant value or prescription to model the diverse population of stars within a given galaxy is well established (see e.g. \citealt{jevrabkova2018impact, Zhang18, hopkins2018dawes, martin2021fornax} for IMFs, \citealt{Heger03massive}\footnote{Figure 1. in \citet{Heger03massive} demonstrates the shift that occurs for final fate mass boundaries as a function of $Z$ (e.g. whether a star should end as an Fe CCSNe with a neutron star remnant, undergo a black hole by fallback or evolve directly into a black hole). The lower mass counterpart to this figure is presented in \citet{Doherty15super} (see their Figure. 5).} 
for final fate upper mass bounds and \citealt{chomiuk2011toward, kennicutt2012star} for SFRs.) 
In any case, there is a notable difference between the Fe CCSNe rates calculated using metallicity--dependent $M_{\rm mas}$ values versus those using a fixed (Z-independent) $M_{\rm mas}$. For example, if we compare our curve to the constant SNe rate using $M_{\rm mas}=8$, we discern a 30\%  deviation between the two at $Z \sim 0.04$. Even close to solar metallicity ($Z=0.015$), the deviation is still 23\%. The smallest variation is actually found at $Z=0.10$, where the difference between the two curves is less than 1\%. 

Beyond Fe CCSNe rates, the metallicity dependence of $M_{\rm mas}$ also has implications for theoretical galactic chemical evolution (GCE) simulations (for reviews on the topic of chemical evolution in galaxies, see e.g. \citealt{matteucci1986relative, matteucci1986evolution, matteucci2012chemical, wheeler1989abundance, Kobayashi20, romano2022evolution}). Consider $M_i$, the amount of available gas within a given galaxy comprising the chemical element $i$. For given $i$, we can use an equation for GCE (e.g. their Equation 4.1 in \citealt{matteucci2012chemical}) to calculate $\dot{M}_i(t)$; i.e., the rate of change of $M_i$ within the galaxy. To perform this calculation, we consider how much $M_i$ is lost from the interstellar medium to form stars or through galactic winds, how much $M_i$ is then expelled from those stars and \textit{returned} to the interstellar medium over their lifetimes (via stellar mass loss, binary interactions and SNe), as well as the composition--specific mass $M_i$ gained from gas infall. As we have emphasized in this work, the quantity and chemical composition of the gas expelled from a star over its lifetime is heavily dependent on the initial mass of the star. As such, achieving the correct integral bounds for the terms in Equation 4.1 from \citet{matteucci2012chemical}, which ultimately governs how much $M_i$ is returned to the ISM, is critical. 

\section{Conclusions} 
\label{section:conclusion}

In this study, we have calculated $M_{\rm mas}$, the minimum initial mass required for stars to undergo a Fe CCSNe event, as a function of initial metallicity and presented the first results for super metal-rich models ($Z_{\rm max}=0.10$). We find that for the metallicity range $Z \approx 1\times 10^{-3}$ to $Z\approx0.04$, the impact of increasing $\kappa$ with $Z$ results in lower $T_{\rm eff}$ and $L$ for their entire evolution. Higher initial masses are then required to ignite core C burning and undergo Fe CCSNe. At approximately $Z=0.05$, we find there is a reversal in this trend, where the impact of increasing $\mu$ with $Z$ begins to dominate. These---most metal-rich models---experience greater $T_{\rm eff}$ and $L$ on the main sequence and produce more massive H exhausted cores. Thus, $M_{\rm mas}$ begins to decline here as $Z$ extends to $Z=0.10$. These results rely on the linear scaling of initial He with $Z$. Our results imply that galactic evolution models are under--predicting SNe rates in the most metal--rich regions if using an extrapolation of a metallicity dependent curve (such as \citealt{Ibeling13}). We caution that the use of non-metallicity-dependent approximations do not reflect the sensitivity of $M_{\rm mas}$ to (even small changes in) $Z$. As such, they are inappropriate, particularly for chemical evolution studies of spiral galaxies and giant ellipticals with metal-rich regions. 

\section*{Acknowledgements}

This research was supported by the Australian Research Council Centre of Excellence for All Sky Astrophysics in 3 Dimensions (ASTRO 3D), through project number CE170100013. Our models were run on the OzSTAR national facility at Swinburne University of Technology. The OzSTAR program receives funding in part from the Astronomy National Collaborative Research Infrastructure Strategy (NCRIS) allocation provided by the Australian Government. 
M. Joyce acknowledges the Lasker Data Science Fellowship awarded by the Space Telescope Science Institute, which supported the development of \texttt{inlists} used in this study and Paper I.  
%
M.Joyce gratefully acknowledges funding of MATISSE: \textit{Measuring Ages Through Isochrones, Seismology, and Stellar Evolution}, awarded through the European 
Commission's Widening Fellowship. This project has received funding from the European Union's Horizon 2020 research and innovation programme.
M. Joyce and G. Cinquegrana further acknowledge the \texttt{MESA} developers and \texttt{MESA} organization for open-source practices and knowledge sharing. G. Cinquegrana is grateful to C. Doherty for discussions about second dredge up as well as A. Heger and our referee, R. Hirschi, for their very helpful comments on the manuscript. 
\section*{Data Availability}

\texttt{MESA}, the software used to produce the simulations for this paper, is a fully open-source stellar evolution code available at \url{https://github.com/MESAHub/}. Following the best practices of the \texttt{MESA} community, data and inlists required to reproduce our results are publicly available at \doi{10.5281/zenodo.8237700}.




\bibliographystyle{mnras}
\bibliography{Giulias_bib, more_bib} 



\appendix

\section{Other figures}\label{A:otherfigures}

\begin{figure*} 
    \centering
	\includegraphics[width=18cm]{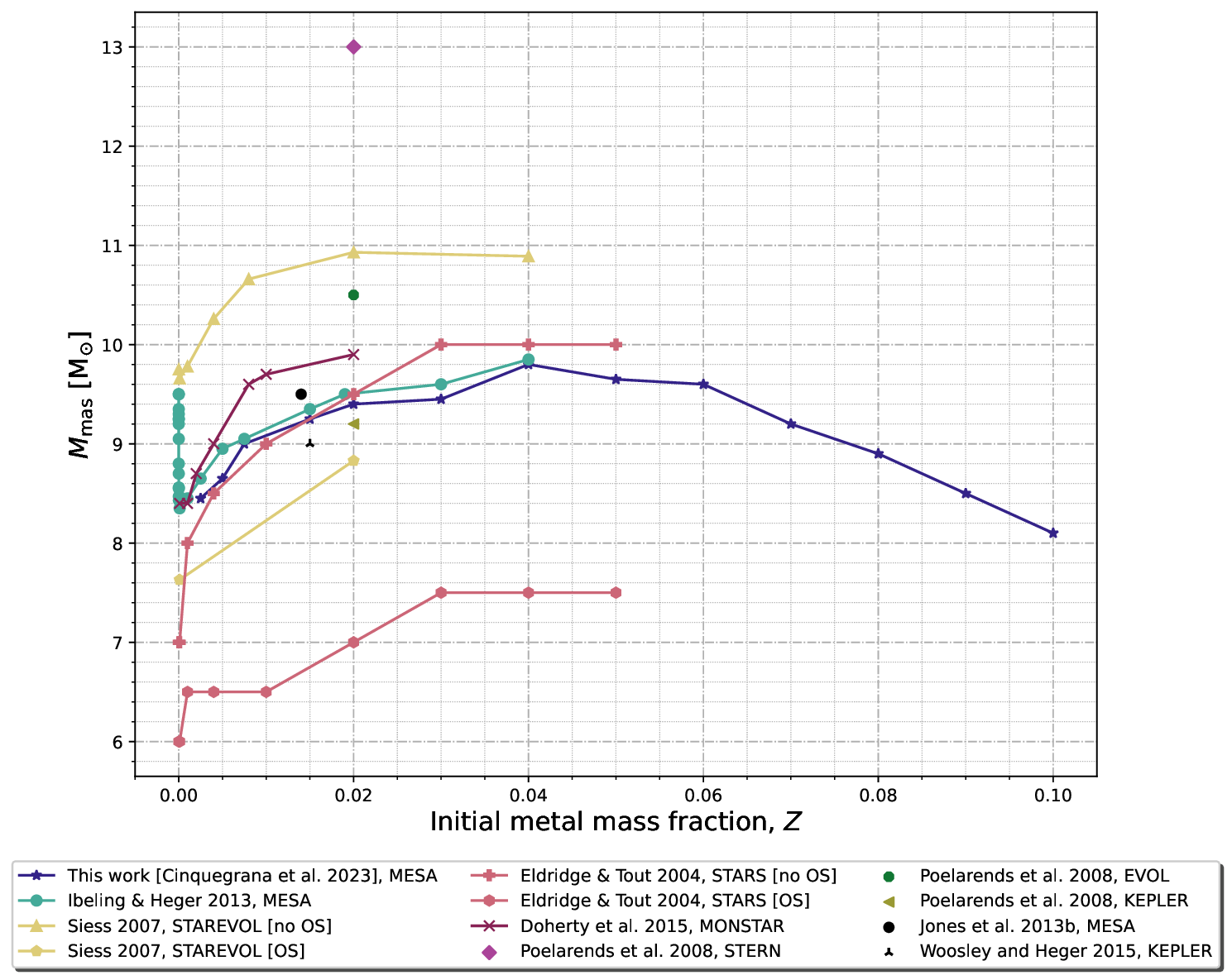}
    \caption{A comparison of our calculated $M_{\rm mas}$ against that currently available in the literature, where $M_{\rm mas}$ is the minimum initial mass required for a star to undergo an Fe CCSNe. The relevant input physics used in each set of models, where available, is listed in Table. \ref{physics_comparison}.}
    \label{fig:append_comp}
\end{figure*}


\bsp	
\label{lastpage}
\end{document}